\documentclass[preprint,times,tighten]{aastex6}
\usepackage{amssymb,amsmath,graphicx,natbib,hyperref,titlesec,url}
\usepackage{color,enumitem}

\renewcommand{\abstractname}{Executive Summary}
\newenvironment{longdescription}
  {\begin{description}[style=unboxed]}
  {\end{description}}

\newcommand{\hi}          {\mbox{\rm H{\small I}}}

\titleformat{\section}[block]
{\normalfont\large\filcenter\sffamily}{}{.5em}{\bfseries}

\titleformat{\subsection}[block]
{\normalfont\sffamily}{}{.5em}{\itshape}

\shorttitle{Key Science Goals for the ngVLA}
\shortauthors{ngVLA Science Advisory Council}

\begin{document}

\begin{flushright}
{\bf ngVLA Memo \#125}
\end{flushright}

\title{Key Science Goals for the Next Generation Very Large Array (ngVLA):\\
Update from the ngVLA Science Advisory Council (2024)}


\author{
David J. Wilner${}^{1}$, 
Brenda C. Matthews${}^{2}$, 
Brett McGuire${}^{3}$,
Jennifer Bergner${}^{4}$,
Fabian Walter${}^{5}$,
Rachel Somerville${}^{6}$,
Megan DeCesar${}^{7}$,
Alexander van der Horst${}^{8}$,
Rachel Osten${}^{9}$,
Alessandra Corsi${}^{10}$,
Andrew Baker${}^{11}$, 
Edwin Bergin${}^{12}$, 
Alberto Bolatto${}^{13}$,
Laura Blecha${}^{14}$,
Geoff Bower${}^{15}$,
Sarah Burke-Spolaor${}^{16}$,
Carlos Carrasco-Gonzalez${}^{17}$,
Katherine de Keller${}^{18}$,
Imke de Pater${}^{4}$,
Mark Dickinson${}^{19}$,
Maria Drout${}^{20}$,
Gregg Hallinan${}^{18}$,
Bunyo Hatsukade${}^{21}$,
Andrea Isella${}^{22}$,
Takuma Izumi${}^{23}$,
Megan Johnson${}^{24}$,
Joseph Lazio${}^{25}$,
Adam Leroy${}^{26}$,
Thomas Maccarone${}^{10}$,
Betsy Mills${}^{27}$,
Munetake Momose${}^{28}$,
Cherry Ng${}^{29}$,
Eric Rosolowsky${}^{30}$,
Nami Sakai${}^{31}$,
Anton Zensus${}^{32}$ \\
\bigskip
{\footnotesize\textnormal{%
${}^{1}${Center for Astrophysics $|$ Harvard \& Smithsonian},
${}^{2}${National Research Council of Canada}, 
${}^{3}${MIT}, 
${}^{4}${Univ.~California, Berkeley}, 
${}^{5}${Max Planck Institut f\"ur Astronomie}, 
${}^{6}${Flatiron Instiute}, 
${}^{7}${George Mason Univ.}, 
${}^{8}${George Washington~Univ.}, 
${}^{9}${Space Telescope Science Institute},
${}^{10}${Texas Tech Univ.},
${}^{11}${Rutgers Univ.}, 
${}^{12}${Univ.~Michigan}, 
${}^{13}${Univ.~Maryland}, 
${}^{14}${Univ.~Florida}, 
${}^{15}${Academia Sinica Institute of Astronomy and Astrophysics}, 
${}^{16}${West Virginia Univ.}, 
${}^{17}${UNAM-IRyA}, 
${}^{18}${California Institute of Technology}, 
${}^{19}${NOIRLab}, 
${}^{20}${Univ. Toronto}, 
${}^{21}${Univ. Tokyo}, 
${}^{22}${Rice Univ.}, 
${}^{23}${NAOJ}, 
${}^{24}${USNO}, 
${}^{25}${JPL}, 
${}^{26}${The Ohio State~Univ}, 
${}^{27}${Univ.~Kansas}, 
${}^{28}${Ibaraki Univ.}, 
${}^{29}${SETI Institute}, 
${}^{30}${Univ.~Alberta}, 
${}^{31}${Institue of Physical and Chemical Research (RIKEN)}, 
${}^{32}${Max Planck Institut f\"ur Radioastronomie} 
    }}}

\maketitle

\section*{Executive Summary \\}

In 2017, the next generation Very Large Array (ngVLA) Science Advisory Council, together with the 
international astronomy community, developed a set of five Key Science Goals (KSGs) to inform, 
prioritize and refine the technical capabilities of a future radio telescope array for high angular 
resolution operation from 1.2-116~GHz with $10\times$ the sensitivity of the Jansky VLA \& ALMA 
\citep[ngVLA memo \#19,][]{ngVLA19}. 
The resulting five Key Science Goals, which require observations at centimeter and millimeter
wavelengths that cannot be achieved by any other facility, represent a small subset of the 
broad range of astrophysical problems that the ngVLA will be able address. 
On account of its scientific importance and versatility, the Astro2020 Decadal Survey on 
Astronomy and Astrophysics of the U.S. National Academies of Sciences endorsed the ngVLA as 
a high priority for new ground-based observatories for construction in the next decade 
\citep{2021pdaa.book.....N}.

This document presents an update to the original ngVLA Key Science Goals, 
taking account of  new results and progress in the $7+$ years since their initial presentation, 
again drawing on the expertise of the ngVLA Science Advisory Council and the broader community
involved in the ngVLA Science Working Groups.
As the design of the ngVLA has also matured substantially in this period
\citep[e.g,][]{ngVLA_systems_reference_design}, this document also briefly 
addresses initial expectations for ngVLA data products and processing that will be 
needed to achieve each of the Key Science Goals (summarized in the Appendix). 
The ngVLA Key Science Goals span the three major Science Themes of Astro2020: 
{\em Worlds and Suns in Context} (KSG1, KSG2), 
{\em New Messengers and New Physics} (KSG4, KSG5), and 
{\em Cosmic Ecosystems} (KSG3). 
As elaborated on in detail in this document, the original five ngVLA Key Science Goals endure as 
outstanding problems of high priority for the field. In many cases, the essential contributions of 
the ngVLA will be bolstered by multi-wavelength/multi-messenger synergies with other new, 
large-scale facilities that are anticipated to be constructed and become available on a similar timeline.
In brief, the ngVLA Key Science Goals are: \\

\begin{longdescription}
\item[Unveiling the Formation of Solar System Analogues (KSG1)]%
The ngVLA will probe the formation of planetary systems similar to our own by revealing structure
in millimeter dust continuum emission at scales of terrestrial planets around large samples of
young stars. Detailed simulations of planet-disk interactions show  that ngVLA observations can 
detect gaps produced by a forming Earth mass planets around a Sun-like stars in nearby dark clouds. 
For gas giant planets, the ngVLA will also detect the azimuthal structures predicted to form in 
disks of low viscosity. Multi-epoch observations over orbital timescales will yield  ``movies'' 
that offer new ways to probe the physics of planet-disk interactions. Population studies will constrain 
the planet initial mass function and provide a framework to understand the origins of exoplanet demographics. 

\item[Probing the Initial Conditions for Planetary Systems and Life with Astrochemistry (KSG2)]%
The ngVLA will be able to detect predicted, but as yet often unobserved in a variety of environments, complex prebiotic species that are the basis of our understanding of chemical evolution toward amino acids and other biogenic molecules.  The sensitivity and resolution of the ngVLA will enable these detections at all stages of the star- and planet-forming cycle, from the earliest pre-stellar cores, through the formation of protostars and protoplanetary disks, to the comae of comets in our own solar system.  Beyond even our own galaxy, the ngVLA will enable the study of this chemistry on entire cloud scales by observations of external galaxies.  The sensitivity of the ngVLA will further permit the study of rare elements and isotopologues, providing fundamental insight into the chemical evolutionary pathways evolving alongside, and shaping, the star- and planet-forming process. The detection of such species will provide the chemical initial conditions of forming solar systems and individual planets.   

\item[Charting the Assembly, Structure, and Evolution of Galaxies from the First Billion Years to the Present (KSG3)]
With its unprecedented collecting area and spatial resolution, the ngVLA will transform our ability to study the physics of the molecular gas phase in galaxies, both in the local universe and at high redshift.
The ngVLA will provide an order-of-magnitude improvement in depth and area for surveys of cold gas in galaxies back to early cosmic epochs, and it will enable routine sub-kiloparsec scale resolution imaging of gas reservoirs, thus constraining their dark matter content. 
In the local universe, the ngVLA will image atomic gas in extended galactic disks and the circumgalactic medium, and survey the physical and chemical properties of molecular gas over the entire local galaxy population. In this way, ngVLA will constrain how galaxies accrete and expel gas, and how they process it into stars. These studies will shed light on the detailed workings of the physical processes that shape galaxy assembly and evolution throughout most of the history of the universe, providing critical constraints for theoretical models and simulations.

\item[Science at the Extremes: Pulsars as Laboratories for Fundamental Physics]

The ngVLA will provide new insights into fundamental physics, primarily using pulsars in the Galactic Center that are predicted to become detectable with the ngVLA. The single highest-reward science goal is to 
find and time one or more pulsars orbiting Sgr A* in order to place extremely stringent constraints on the
supermassive black hole's properties and test for deviations from general relativity. 
Additional fundamental science includes gravitational tests using pulsar-black hole binaries; narrowing in
on the dense matter equation of state with pulsar binaries, especially when combined with next-generation 
multi-wavelength and multi-messenger observatories; determining the origin of the Galactic Center GeV excess, 
in particular whether it is dark matter annihilation or magnetospheric emission from millisecond pulsars; 
and probing the astrophysics of low-frequency gravitational wave sources. The ngVLA's combination of 
high frequencies, high sensitivity,  
and long baselines will drastically improve the outcomes of these studies; it is likely the only 
facility on the horizon capable of achieving most of these fundamental physics goals.

\item[Understanding the Formation and Evolution of Stellar and Supermassive Black Holes in the Era of Multi-Messenger Astronomy (KSG5)]
The ngVLA will be the ultimate black hole hunting machine, surveying everything from the remnants of massive stars to the supermassive black holes that lurk in the centers of galaxies, while providing a breakthrough inventory of intermediate mass black holes in hundreds of globular cluster systems. High-resolution imaging abilities will allow us to separate low-luminosity black holes in our local Universe from background sources, thereby providing critical constraints on the formation and growth of black holes of all sizes and mergers of black hole-black hole binaries. The ngVLA will also identify the radio counterparts to transient sources discovered by gravitational wave, neutrino, and optical observatories. Its high-resolution, fast-mapping capabilities will make it the preferred instrument to pinpoint transients associated with violent phenomena such as supermassive black hole mergers and blast waves.
\end{longdescription}

\clearpage

\section*{Introduction \\}

The next generation Very Large Array (ngVLA) is a radio telescope that will improve on the sensitivity 
and angular resolution of the Jansky VLA and ALMA by more than an order of magnitude at frequencies 
from 1.2--116 GHz. The ngVLA is envisioned to consist of approximately 
244 antennas of 18 meter diameter configured with (1) a dense core within $\sim1$~km that also 
includes 19 close-packed antennas of 6 meter diameter for short spacings, 
(2) a main array that reaches $\sim1000$~km baselines, 
and also (3) extended baselines to continental scales.
On account of its scientific importance and versatility, the Astro2020 Decadal Survey 
on Astronomy and Astrophysics of endorsed the ngVLA as a high priority for new ground-based observatories  for construction in the next decade \citep{2021pdaa.book.....N}.

Through a process begun in 2017, ``Key Science Goals'' (KSGs) for the ngVLA were developed 
by the ngVLA Science Advisory Council from a large number of science use cases that were 
submitted by the international community. 
These use cases outline science drivers together with the technical specifications required of 
a facility to achieve the science goals described. 
New science use cases continue to be submitted.
To identify Key Science Goals from among the many use cases 
\citep[summarized in the ngVLA Science Book,][]{Murphy2018}, the 
Science Advisory Council assessed all of them under the following criteria: 

\begin{itemize}
    \item{the compelling nature of the science question;}
    \item{the unique ability of the ngVLA to address the science question;}
    \item{the ability of ngVLA to open up a new discovery space and/or deliver a breakthrough discovery; and}
    \item{the complementarity with existing and future facilities to address the science question.}
\end{itemize}

\noindent
The resulting five ngVLA Key Science Goals, which require observations at centimeter and millimeter 
wavelengths that cannot be achieved by any other facility, were summarized in ngVLA memo \#19 \citep{ngVLA19}. 
These Key Science Goals have been revisited, and, in no particular order, they are:

\begin{enumerate}
\item Unveiling the Formation of Solar System Analogues
\item Probing the Initial Conditions for Planetary Systems and Life with Astrochemistry 
\item Charting the Assembly, Structure, and Evolution of Galaxies from the First Billion Years to the Present: 
\item Science at the Extremes: Pulsars as Laboratories for Fundamental Physics
\item Understanding the Formation and Evolution of Stellar and Supermassive Black Holes in the Era of Multi-Messenger
Astronomy
\end{enumerate}

\noindent
Notably, these ngVLA Key Science Goals align closely with the three major Science Themes (and
priority areas) for the next decade subsequently identified by the Astro2020 Decadal Survey: 
{\em Worlds and Suns in Context (Pathways to Habitable Worlds)} (KSG1, KSG2), 
{\em New Messengers and New Physics (New Windows on the Dynamic Universe)} (KSG4, KSG5), and 
{\em Cosmic Ecosystems (Unveiling the Drivers of Galaxy Growth)} (KSG3). 
The Astro2020 panel on Radio, Millimeter, and Submillimeter Observations 
from the Ground details the importance of ngVLA capabilities for addressing the specific
high-priority science questions and discovery areas identified by the Decadal Survey 
science frontier panels (see Appendix M of Astro2020 report).

In the $7+$ years since the initial formulation of these Key Science Goals, observations with 
new (and old) facilities have produced many new results, theory and computation continue 
to improve, and astrophysical understanding has advanced considerably. 
Given these ongoing developments, the ngVLA Science Advisory Council, together with the 
broader community represented by the ngVLA Science Working Groups, has carefully re-examined 
the ngVLA Key Science Goals, keeping with the original criteria for their evaluation and selection.
As detailed below, these five ngVLA Key Science Goals endure as outstanding problems 
of high priority, as essential aspects {\em require} the ngVLA to address. 

This document presents an update to the five ngVLA Key Science Goals, 
taking account applicable progress in the relevant science areas, in the sections that follow.
An appendix to this document briefly summarizes initial expectations for ngVLA data deliverables 
needed to realize each of these Key Science Goals, in light of the more mature ngVLA design 
\citep{ngVLA_systems_reference_design} and the observing modes and data products 
\citep{observing_modes_calibration_strategy} now expected to be available.

\clearpage

\section{1. Key Science Goal: Unveiling the Formation of Solar System Analogues \\}

\subsection{Scientific Rationale}
Planets are assembled in disks of gas and dust orbiting pre-main sequence stars, but the physical 
processes responsible for  their formation and the variables that account for their diversity remain 
to be established. The fundamental properties of planets, including their masses, compositions, and 
orbital radii, are apparently largely set during an early formation epoch spanning less than 10~Myr. 
The angular resolution required to spatially resolve the planet-forming regions of nearby disks 
has been achieved only in the last decade with ALMA and the largest optical telescopes, and these observations 
have resulted in the discovery that many disks exhibit  morphological features such as rings, spirals and 
arcs suggestive of gravitational perturbations by unseen giant planets \citep[see the review by][]{Andrews2020}.

Characterization of disk substructures provides a powerful tool to estimate masses and orbital radii 
of protoplanets, investigate circumplanetary environments, and determine how forming planets interact 
with circumstellar material \citep[e.g.][]{vanderMarel2021,Bae2022,Manara2022}. 
For one young planetary system, PDS 70, ALMA observations have detected the
{\em circumplanetary disks}  \citep{Isella2019,Benisty2021} around two giant planets 
in formation at orbital radii of 22 au and 34 au 
\citep{Keppler2018,Muller2018,Haffert2019}
that have carved a deep cavity in the {\em circumstellar disk}, 
highlighting its capability to detect signatures of protoplanets on long period orbits.  
These protoplanets, PDS 70b and PDS 70c, are estimated to exceed Jupiter's mass by a factor of a few 
and formed very fast at very early stages of the disk evolution. 
The less massive, terrestrial planets are believed to form via slower grain growth within substructures 
and/or in the innermost ($<10$~au) regions of the disks. 
While these are compact and very dense regions 
with highly opaque emission at wavelengths shorter than a few millimeters, 
their study have been proved possible through observations at longer wavelengths 
\citep{Carrasco-Gonzalez2019}.
{\bf The ngVLA, with higher resolution and sensitivity at millimeter-to-centimeter wavelengths, 
will open a window into the terrestrial planet-forming regions of nearby disks, 
within a few au of the central star, enabling the 
detection of dusty substructures associated with less massive planets on shorter-period orbits.} 
The high resolution of ngVLA will probe and track the formation of 
super-Earths and sub-Neptunes, which appear to be the most numerous planets in the Galaxy.

The population of young planets in inner disks accessible to the ngVLA will overlap with the 
thousands of mature planets that have been detected by various  techniques orbiting main-sequence stars. 
In particular, giant exoplanets are known to be most common at orbital radii of 2-8 au \citep{Fulton2021}, 
well within reach of the ngVLA. Recent modeling by \citep{Zhu2023} shows that at centimeter wavelengths, 
one can see through the optically thin circumstellar
disk and detect the denser envelope within the Hill sphere of an embedded giant planet. 
The diverse chemistry of the active surfaces of these inner disk regions is being revealed through 
mid-infrared spectroscopy, particularly with JWST \citep{Pontoppidan2024}, though the underlying 
structures are not spatially resolved. 
High resolution imaging with the ngVLA will revolutionize the study of the birth of exoplanets, providing an 
essential complement to the planned extremely large optical telescopes that are expected to directly 
image exoplanets around nearby stars, including some in habitable zones. 
The new observations of planet-forming disks enabled by the ngVLA align with the Astro2020 
priority area {\em Pathways to Habitable Worlds} within the Science Theme {\em Worlds and Suns in Context}. 

\begin{figure}[t!]
\centering
\includegraphics[width=0.9\textwidth]{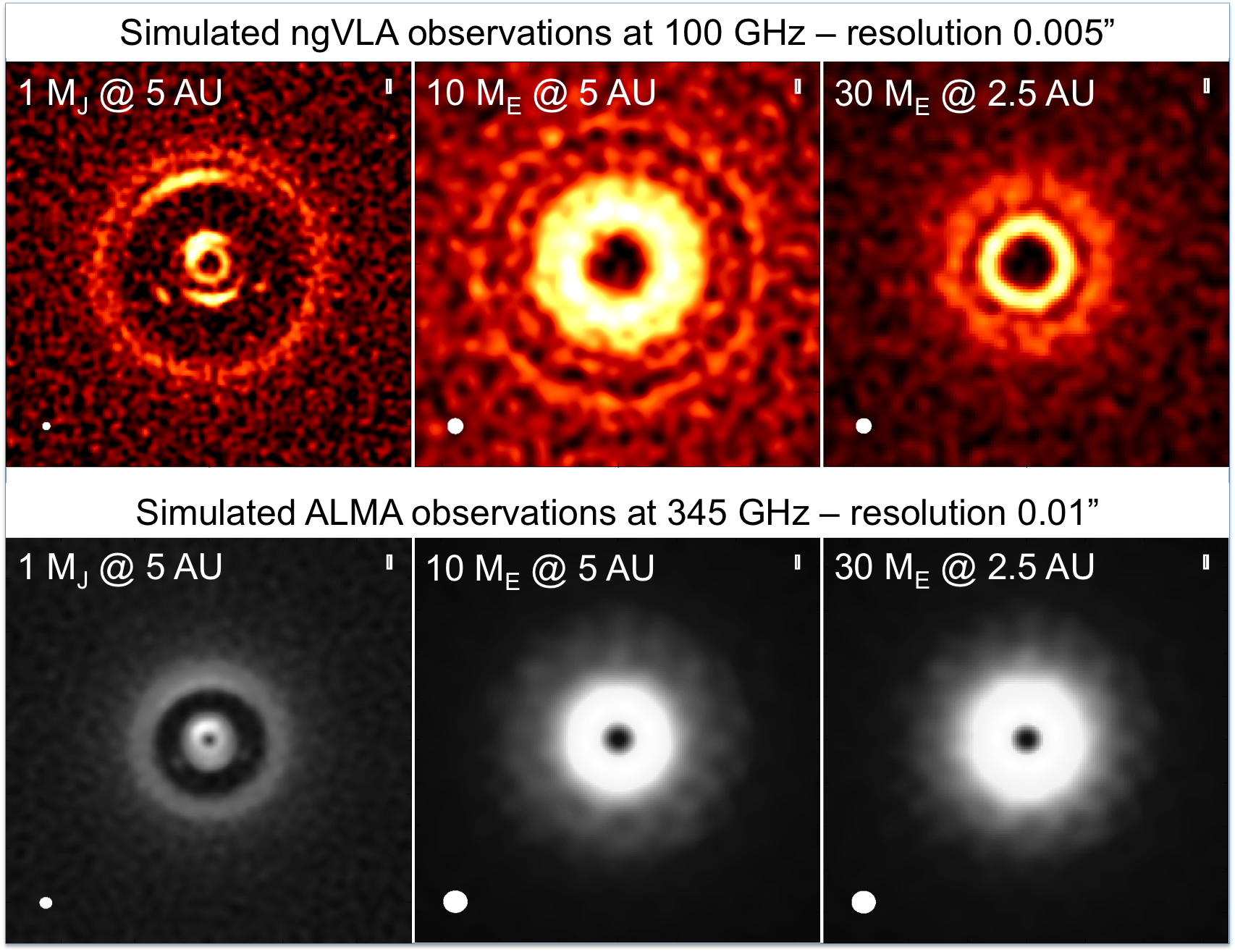}
\caption{Simulated ngVLA observations (top row) and ALMA observations (bottom row) of dust continuum emission 
from a protoplanetary disk perturbed by a Jupiter mass planet orbiting at 5 au (left), 
a 10 Earth mass planet orbiting at 5 au (center), and a 30 Earth-mass planet orbiting at 2.5 au (right). 
The ngVLA observations at 100 GHz were simulated for an angular resolution of 5 mas and a rms noise level 
of 0.5~$\mu$Jy beam$^{-1}$; 
the ALMA observations at 345 GHz were simulated to obtain an angular resolution of 10 mas, 
using observations in its most extended array configuration with baselines extending to 16 km and
a rms noise level of 8~$\mu$Jy beam$^{-1}$ 
\citep{Ricci2018a,Ricci2018b}. 
}
\label{fig:ngVLA-ALMA-disks}
\end{figure}

\subsection{Planet-Disk Interactions at Scales Down to $<1$~au}

Figure~\ref{fig:ngVLA-ALMA-disks} illustrates the capabilities of the ngVLA for imaging planetary systems in the 
act of forming, comparing simulated ngVLA observations at a frequency of 100 GHz to simulated ALMA observations 
at a frequency of 345 GHz, which provide the best compromise between angular resolution and sensitivity to the 
dust thermal emission \citep{Ricci2018a,Ricci2018b}. 
Observations with the ngVLA will clearly reveal the presence of planets with masses as low 
as 10 Earth masses at orbital radii as small as 2.5 au (central and right panels). 
These planets will never be detectable by ALMA because of the high optical depth of the dust emission at 345 GHz 
and its lower angular resolution. The ngVLA will be also superior to ALMA at imaging perturbations generated 
by giant planets orbiting close to the central star (left panels). In particular, localized dust heated by the 
young Jupiter analog is detected as circumplanetary emission within the prominent gap carved in the disk, 
and features located 60 degrees ahead and behind the planet result from dust grains trapped near the L4 and L5 Lagrange points (``Trojans''). 
The exquisite angular resolution of the  ngVLA will enable the measurement of 
the orbital motions of all these structures on monthly timescales to make multi-epoch 
``movies'' of the observed structures \citep{Wilner2004}.
These synoptic observations will add a completely new dimension to the study of giant planet formation. 

\begin{figure}[t]
\centering
\includegraphics[width=0.9\textwidth]{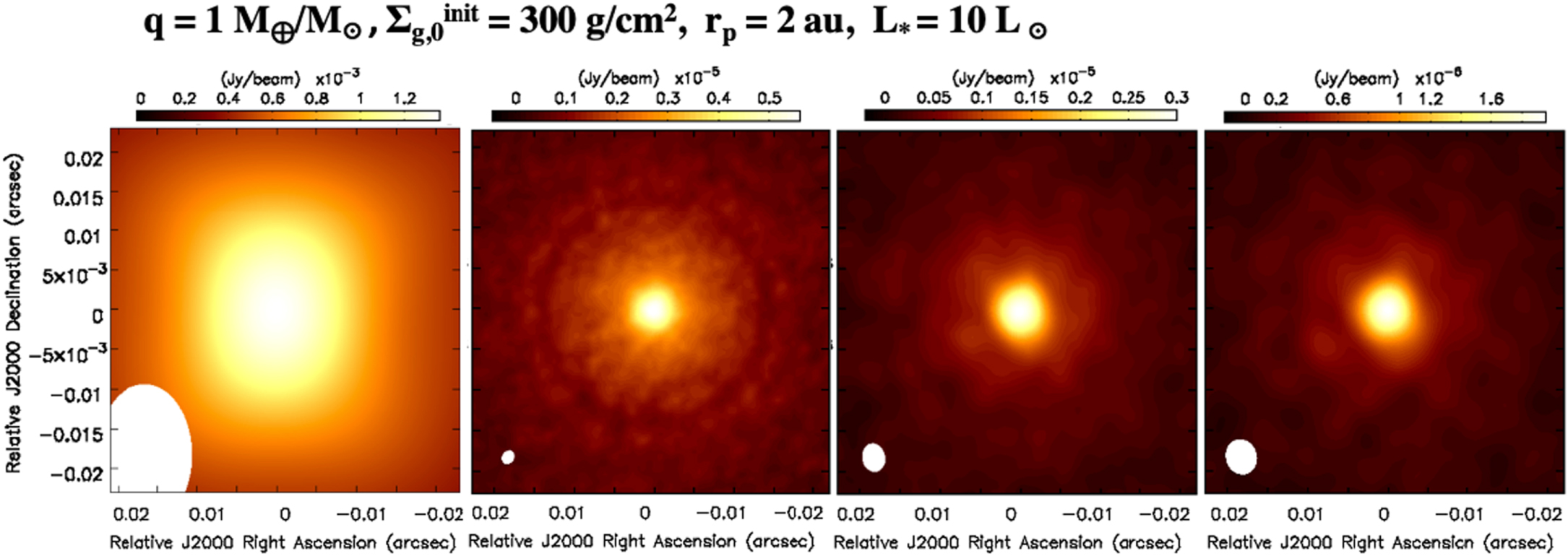}
\caption{
Simulated ALMA observations at 230~GHz (left) and ngVLA observations at 100, 43 and 30~GHz (right) 
of dust continuum emission from a protoplanetary disk perturbed by an Earth-mass planet orbiting at 3 au
around a Solar-mass star \citep{Harter2020}. The synthesized beam sizes are 18~mas (2.5~au), 1.5~mas (0.2~au),
3 mas (0.4~au) and 4 mas (0.55~au), respectively, with rms noise values 2800, 86, 35, and 21 nJy~beam$^{-1}$. 
The ngVLA angular resolution is essential for discerning the gap opened by the planet in the disk.
} 
\label{fig:ngVLA-ALMA-Earthlike}
\end{figure}

The angular resolution and sensitivity of multi-wavelength disk imagery is currently 
limited to probing for the presence of planets more massive than Neptune at orbital 
radii beyond $10$~au. A key science goal of the ngVLA will be to image the formation of 
super-Earths and giant planets across the entire radial distribution of the disk, particularly 
at $1 -10$~au from the central star, and to probe for the presence of planets with masses 
as low as 1 Earth mass, which could be harboured even in disks currently thought not to exhibit substructure.  
The comparison of the occurrence of protoplanets at large orbital radii with exoplanet populations will 
help to disentangle the question regarding in situ formation vs formation followed by inward migration, 
which is a key issue in current planet formation models.
Numerical hydrodynamical simulations of the combined evolution of gas and dust in disks with 
embedded planets indicate that ngVLA will be able to detect signatures of Earth-mass planets in 
Earth-like orbits around Sun-like pre-main-sequence stars \citep{Harter2020}. 
Figure~\ref{fig:ngVLA-ALMA-Earthlike} shows simulated observations for the dust continuum emission 
at 230 GHz (ALMA) and 100, 43, and 30 GHz (ngVLA) for a disk model with a planet with planet-to-star mass 
ratio of 1~M$_{\oplus}$/M$_{\odot}$ at 3~au from the host star, showing that ngVLA can reveal the 
subtle gap created by the planet. 

\subsection{Constraints on the Planet Initial Mass Function}

The indirect detections of large numbers of exoplanets in formation will provide strong constraints on  
the initial mass function and the birth orbital radius function of giant and massive rocky planets. 
By observing dust substructures resulting from planet-disk interactions, the ngVLA will open up new territory 
in the parameter space of planet mass - orbital radius for young planets, with the potential to reveal of 
hundreds of planets with masses $> 5$~M$_{\oplus}$ at orbital radii $< 10$~au in regions out to 1~kpc distance.
These indirect observations of young planets, prior to any potential subsequent dynamical rearrangement, 
will 
provide key information to understand the origin and diverse demographics of exoplanetary systems, and, ultimately, 
unveil the formation of planetary systems similar to our own Solar system.
Our Jupiter, for example, could have prevented inward migration of super-Earth cores that likely formed 
in the Solar System outer regions, thereby protecting innner rocky planets 
from dynamical disruption \citep{Izidoro2015a}. The presence of Jupiter may then explain the lack 
of otherwise extremely common Super-Earths in the Solar System \citep{Batygin2015}, as well as the 
formation of Uranus and Neptune \citep{Izidoro2015b}.
Comparisons of simulated observations from multifluid hydrodynamic numerical simulations 
for ngVLA, SKA-mid (under construction in South Africa), and the proposed SKA-2, show that 
only ngVLA -- operating at subcentimeter wavelengths -- can produce high-fidelity images  
of the relevant disk substructures in reasonable integration times 
to enable surveys \citep{Wu2024}. 

\subsection{Grain Growth and Dynamics}

Grain growth is expected to play a key role in the formation of planetesimals and planetary embryos. 
At $10-50$ GHz the ngVLA will image 
emission from dust grains with characteristic size $\sim1$~cm, the detailed distribution
of which 
is inaccessible to ALMA and the VLA, 
spatially separating emission from disk dust any ionized gas in outflows that may be present
at these frequencies \citep{Tobin2018}.
Together with optical/infrared scattered light observations that probe smaller dust grains,
of order $\sim1$~$\mu$m in size, these radio wavelength observations 
of emission from millimeter-to-centimeter size dust grains will provide 
information on grain growth, vertical settling, and radial drift, to probe the dynamical 
interaction between dust and gas particles 
before, during, and after planet formation. Where signal-to-noise is sufficient, full polarization observations 
can provide constraints on dust grain properties and potentially magnetic field morphologies. 

\subsection{Ionized Flows}
The ngVLA's angular resolution will be sufficient to detect and distinguish ionized gas emission 
from a wind or jet from dust emission from the disk. Such observations will provide new constraints 
on the physical mechanism that drives ionized flows, where present. These flows are fundamental processes 
for disk evolution: photoevaporative winds are likely important to disk dispersal
\citep{Pascucci2022}, while magnetohydrodynamic winds are gaining popularity as the agent for angular 
momentum transport that enables disk accretion \citep[e.g.][]{Lesur2022}. 
\citet{Ricci2021} present simulations of free-free emission at 1~cm wavelength obtained from 
models of photoevaporative and magnetohydrodynamic winds from a T-Tauri star; these different mechanisms 
show different morphologies that ngVLA imaging observations can clearly discriminate. 

\subsection{Spectral Lines: Disk Kinematics}
In addition to continuum observations of thermal dust emission, spectral line emission from protoplanetary 
disks will be essential to probe the kinematics in planet-forming disks. 
Perturbations of gas motions from Keplerian rotation are particularly important to link dust substructures 
to protoplanets, as the dust features such as gaps and rings may arise from other mechanisms, 
such as fluid instabilities \citep{Lesur2022} or concentrations at the edge of  ``dead zones''  
\citep{Pinilla2016,Ueda2022} that have different kinematic signatures.
Synthetic channel maps of ngVLA observations of $^{12}$CO 1-0 emission derived from disk models show clearly the 
kinematic ``kink'' due to gravitational perturbations to the Keplerian velocity field by a Jupiter-mass planet 
at an orbital radius of 10's of au \citep{Ricci2022-ngVLA101}.
While some redundancy exists with species detectable in the ALMA lowest frequency bands, 
bulk gas tracers like CO isotopologues and the low excitation lines of other important trace species
will be observable with an order of magnitude more sensitivity than ALMA, enabling higher resolution.

\subsection{Telescope Requirements}
Continuum imaging from 20--115 GHz with an angular resolution of 1.5~mas is needed to 
detect the signatures of terrestrial mass planets due to planet-disk interactions in the innermost 
$1 -10$~au of protoplanetary disks associated with nearby star forming regions ($\sim$150 pc). 
Detailed numerical simulations of disks perturbed by planets \citep[e.g.][]{Ricci2018a,Harter2020}
show that an rms continuum sensitivity at 100~GHz of 0.5 $\mu$Jy beam$^{-1}$ will be sufficient
to reveal structures in the dust distribution created by planets of interest, in particular
an Earth-mass planet orbiting around a Sun-like pre-main-sequence star in the terrestrial zone.
These observations will benefit from the largest possible aggregate bandwidth, and from full polarization 
capabilities to constrain the properties of the dust grains during the earliest phases of aggregation. 
To image the full extent of CO J=1-0 line emission from the largest disks in a single pointing, 
a field of view at least 10\arcsec\ is required, with comparable maximum recoverable scale.
Simulations show that detection of kinematic perturbations due to giant planets in formation 
through line observations demands spectral resolution of 0.5~km~s$^{-1}$ per channel or higher. 

\subsection{Data Product and Processing Requirements:}
In general, the high level data products made available to users by the ngVLA are likely to be sufficient for 
a significant fraction of protoplanetary disk science. However, because of the trade-offs in sensitivity 
and angular resolution at modest signal-to-noise, for example for continuum imaging disk structures in
the terrestrial zone whose scales and forms are not known  {\it a priori}, we anticipate requiring the
option to reprocess the visibility data with different weighting schemes to create a variety of images 
that highlight different features. To produce confidence in subtle features, experience with ALMA suggests 
several realizations with different imaging parameters may be required.
We also would like access to fully calibrated visibilities in order for direct model-fitting, ideally
pre-averaged in time and frequency as appropriate to the individual sources of that span only a few arcseconds. 
Access to the visibilities would also enable application of imaging and deconvolution algorithms that
are not implemented in the standard software, e.g. for exploring super-resolution or systematic
uncertainties. For spectral line imaging, we would like to preserve information at the native channel spacing 
with the option to select out lines of interest as well as search for serendipitous detections of other lines. 

\bigskip\bigskip\bigskip


\section{2. Key Science Goal: Probing the Initial Conditions for Planetary Systems and Life with Astrochemistry \\}



\subsection{Scientific Rationale}

One of the most challenging aspects in understanding the origin and evolution of planets and planetary systems is tracing the mutual influence of chemistry and physics during the evolution from molecular clouds to solar systems.  The ngVLA will enable unprecedented observations of interstellar chemistry from the densest star-forming regions of the galaxy to protoplanetary disks. Existing facilities have already shown the stunning degree of molecular complexity present in these systems. The unique combination of sensitivity and spatial resolution offered by the ngVLA will permit the observation of both highly complex and very low-abundance chemical species that are exquisitely sensitive to the physical conditions and evolutionary history of their sources, which are out of reach of current observatories. In turn, by understanding the chemical evolution of these complex molecules, heretofore unattainable detailed astrophysical insight can be gleaned from these astrochemical observations.

\subsection{Complex Chemistry at Early Stages of Star-Formation}

The ngVLA will enable an unprecedented view into complex organic (prebiotic) chemical evolution in the ISM. Existing facilities continue to provide a trickle of tantalizing new prebiotic molecular detections, such as the recent discovery of the biomolecule ethanolamine in G+0.693-0.027 \citep{Rivilla:2022:829288}. Observation of a substantial number of predicted, but as yet undetected, complex prebiotic species are needed to truly understand chemical evolution toward amino acids and other biogenic molecules. We are rapidly approaching the point of diminishing returns at which deep observations with ALMA and the GBT will no longer reveal new spectral lines, due to a combination of sensitivity limits and line-confusion at higher frequencies. Both problems can be solved by sensitive observations in the cm-wave regime. State-of-the-art models predict these molecules will display emission lines with intensities that are easily detectable with the ngVLA, but well below the current detectability thresholds of existing telescopes including ALMA, GBT, and IRAM. N, O, and S-bearing small aromatic molecules, direct amino acid precursors, biogenic species such as sugars, chiral molecules, and, possibly amino acids themselves are all potential candidates for discovery with the ngVLA.  

\begin{figure}[bh!]
    \centering
    \includegraphics[width=0.7\textwidth]{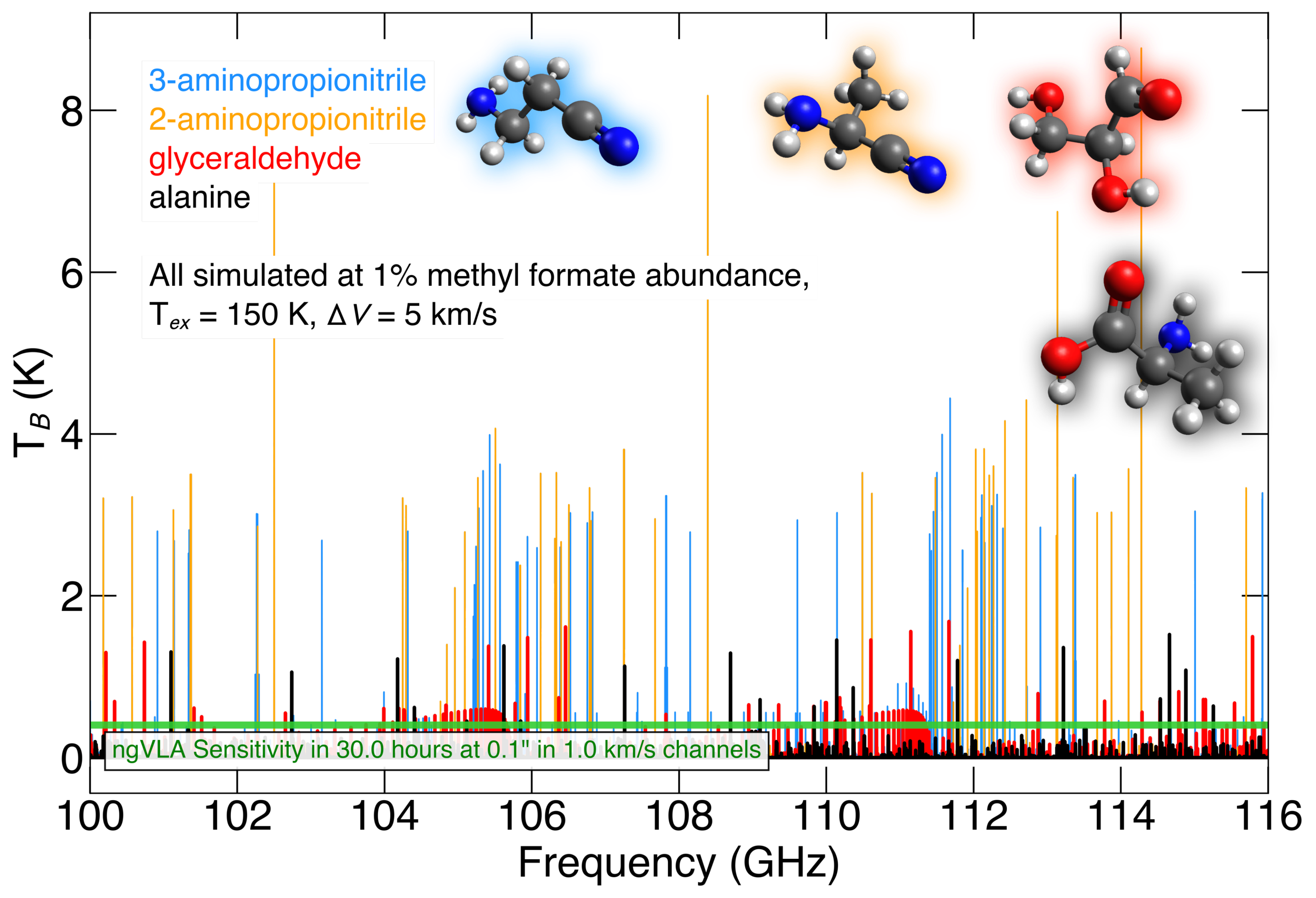}
    \caption{Simulation of emission from four complex organic molecules not yet detected in the ISM but potentially within reach of the ngVLA.}
    \label{ksg2_fig}
\end{figure}

Figure~\ref{ksg2_fig} shows a fiducial simulation of the kinds of molecular complexity which will be detectable with the ngVLA. The molecules were simulated at a column density $\sim$1\% that of methyl formate -- one of the most common indicators of complex organic chemistry -- toward IRAS 16293. The physical conditions were chosen to be somewhat average for star-forming regions.  Many of these parameters -- abundance, temperature, linewidth -- could vary by an order of magnitude and still result in detectable signal within a reasonable integration time for a deep-drill observation.

\subsection{Small-scale protostellar \& protoplanetary environments}
Planet formation takes place in the circumstellar disks surrounding young stellar objects. Circumstellar disks can appear early in a protostar's evolution, when the object is still deeply embedded in its parent envelope (Class 0/I). The disk occupies similar spatial scales as other protostellar structures like the base of the outflow and, in some cases, a `hot corino' of warm, organic-rich gas in the inner envelope. To understand the initial steps of planet formation physics and chemistry, it is imperative to disentangle the connection between these different small-scale protostellar structures, and how they influence each other's chemical and physical evolution. ALMA has provided considerable new insights into the chemical environments of protostellar inner envelopes  \citep[e.g.][]{Imai2019,Garufi2021,Codella2021,Bergner2020,Tychoniec2021} and the diversity of outcomes across different systems \citep[e.g.][]{Yang2021,Bouvier2022,Hsu2022}, but still suffers from high dust opacities and line confusion limits at (sub-)millimeter wavelengths, along with insufficient spatial resolution to distinguish the smallest scale ($<$tens of au) protostellar structures. ngVLA will be a revolutionary facility in this area: cm-wave spectral coverage will mitigate issues with dust opacity \citep{deSimone2020} and line confusion \citep{McGuire2022}, while high angular resolution will disentangle emission of the disk, outflow, and inner envelope/hot corino. Together this will provide an unobstructed view of the chemical environment and evolution relevant to the first steps of planet formation.

In more mature systems, Class II or `protoplanetary' disks, the envelope has cleared and planet formation seems to be well underway, as hinted at by the dramatic substructures (rings, gaps, and spirals) seen in the dust \citep{Andrews2018}. In recent years ALMA has transformed our understanding of protoplanetary disk chemistry: it has revealed a wide and unexplained diversity in chemical emission morphologies \citep{Oberg2021, Law2021}; detected mid-sized organic molecules \citep{Oberg2015,Walsh2016,Booth2021}; and uncovered a distinctive pattern of disk atmosphere compositions enriched in C- and N-bearing molecules and depleted in oxygen \citep[e.g.][]{Schwarz2016,Cleeves2018,Miotello2019}. However, even at its highest possible resolution ALMA can only begin to access spatial scales relevant to solar systems ($\sim$10 au), and is often sensitive to emission from warm, elevated layers of the disk rather than the planet-forming midplane. ngVLA will answer key open questions about planet formation chemistry, such as: Do disks exhibit substructuring in their chemical emission morphologies on solar system (1--10 au) scales?  Is the inner disk characterized by the same distinctive abundance patterns seen in the outer disk?  And, how is the molecular layer chemistry connected to the midplane? ngVLA can target low-energy transitions of key gas-phase chemical tracers (and their rare isotopologues) like CO, HCN, C$_2$H, CN, HCO$^+$, and N$_2$H$^+$, and may also detect mid-sized organics like CH$_3$CN, HC$_3$N, and CH$_3$OH in nearby disks.  Excitingly, ngVLA will also provide unique access to NH$_3$, which cannot be detected at millimeter wavelengths but which may be a major nitrogen carrier, providing much-needed constraints on the nitrogen budget in disks. Observations of these small-molecule tracers at high resolution will necessitate the exceptional sensitivity of ngVLA, and will reveal for the first time the composition and distribution of cool gas-phase volatile material on solar system scales.

\subsection{Astrobiology: Comets \& Chirality}

\paragraph{Comets.} Comets often represent the pristine chemical remnants of their natal molecular clouds. Their proximity to Earth offers a rare opportunity to probe these chemical inventories using both \emph{in situ} instruments from probe missions and sensitive ground-based radio astronomy, for which the ngVLA will be uniquely suited. For example, one particular mystery that will be addressed by the ngVLA is the apparent depletion of nitrogen, where cometary nitrogen levels (relative to carbon) are measured to be as much as an order of magnitude lower than solar \citep{Altwegg:2020:533}. Alongside HCN, {NH3} is the dominant reservoir of nitrogen in comets \citep{DelloRusso:2016:301}, but it is frustratingly challenging to observe. The 23\,GHz inversion transitions offer a potentially powerful observational handle, but have so far been observed in only a handful of comets \citep{Hatchell:2005:777} due to low brightness and complex spatial structure. The ngVLA will have the sensitivity, spectral resolving power, and appropriate coverage across a wide range of spatial scales to observe these {NH3} transitions in a far more represented sample of comets.

\paragraph{Probing the Origins of Chirality.} A highlight of the unique prebiotic science that will be made possible by the ngVLA is the study of chirality and its drivers, particularly the origin of homochirality in biological systems. Chiral molecules, that is, molecules whose mirror image is not identical to the original, are central to biological function. Indeed, the mystery of homochirality, nature's use of only one of the mirror images in most biological processes, plays a central role in our quest to understand the origins of life, as well as being considered a nearly unambiguous biomarker. There is no energetic basis for the dominance in life of one handedness of a chiral molecule over another, but rather, a slight excess was likely inherited at some point in the evolutionary process, and amplified by life. Given that material in planetary systems has been shown to be inherited from their parent molecular clouds, an excess of a particular handedness in that cloud may be the spark which drives homochirality in a certain direction. One possible route to generate a chiral excess is through UV-driven photodissociation of chiral molecules by an excess of left or right circularly polarized light. The ability not only to detect, but to image the abundance of chiral species at spatial scales commensurate with observations of circularly polarized light toward star-forming regions would be a giant leap forward. Using known, polarization-dependent photodissociation cross sections from laboratory studies, these observations would enable quantitative estimates of potential UV-driven excess. While such studies are well beyond the capability of existing observatories, they would be achievable with the ngVLA. Chiral molecules, like other complex species detected earlier, are necessarily large, with propylene oxide, the only detected chiral species to date \citep{McGuire:2016:1449}, being perhaps the only example simple enough for detection with existing facilities. The ngVLA will provide the sensitivity and angular resolution required to detect additional, biologically-relevant chiral species, such as glyceraldehyde. 

\subsection{Extragalactic Astrochemistry}

Molecular line observations are critical tools for probing the chemistry and physics of the ISM in our own galaxy, with increasingly complex molecules providing an ever more-refined picture.  Yet, our position in the Galaxy results in observational constraints that make realizing a holistic understanding of large-scale chemical and physical evolution challenging. Observations of molecular species in external galaxies offer an appealing solution, enabling studies on molecular-cloud-level scales, but until quite recently these observations had been limited to only a handful of the most abundant species. Modern astronomical facilities \citep{Sewilo:2018:L19,Qiu:2018:A3} have now enabled the detection and study of a far larger fraction of the interstellar inventory -- around 30\% -- in external galaxies \citep{2021Census}. While substantial strides have been made using, in particular, ALMA \citep{Martin:2021:A46}, we are once again pushing the technological boundaries of what is possible.  The ngVLA will provide the transformational next leap required both to detect the next level of chemical complexity, but also to enable much larger surveys of representative sample sizes of molecular clouds in external galaxies (and indeed, larger samples of external galaxies themselves).

\subsection{Rare elements \& isotopologues}
Across all galactic and extragalactic environments, ngVLA's unprecedented sensitivity will open a new discovery space for studying the chemistry of rare elements and isotopologues.  Molecular astrochemistry is biased towards the study of abundant elements that are commonly present in the gas phase, i.e.~CHNO and, to a lesser extent, S \citep{McGuire2022}. With dedicated effort, emission from molecules containing heavier atoms like P, Na, K, and Cl can be detected \citep{Rivilla2020,Bergner2022,Ginsburg2023}, though these are generally expensive studies even with ALMA. Indeed, heavier atoms generally exhibit rotational transitions at lower frequencies, making sensitive coverage of the cm regime a promising avenue for expanding our understanding of rare-element astrochemistry. Given their low volatility, searches for new rare-element molecules will likely be most fruitful towards shocked star-forming environments. A particularly impactful direction will be increasing the inventory of inorganic and organic S and P carriers detected, providing a more complete understanding of the chemistry and budget of these biogenic elements. Moreover, detecting molecular carriers of other heavier elements for which we have even fewer constraints may ignite a new subfield of astrochemistry involving the chemistry of organometallic compounds and/or the inheritance of refractory planetary building blocks.

The high sensitivity of ngVLA will also advance the study rare HCNO(S) isotopes, which are tens to hundreds of times less abundant than the major isotopes $^{1}$H, $^{12}$C, $^{14}$N, $^{16}$O, and $^{32}$S. The incorporation of rare isotopologues into molecules at non-statistical ratios, or isotope `fractionation', can be driven by distinctive physical conditions like low temperatures or high UV fields. Studying the fractionation of molecule tracers (e.g.~HCN/HC$^{15}$N/DCN) can therefore provide a powerful tool for discerning the physical conditions in which they formed. Moreover, comparing these fractionation patterns along the star- and planet-formation sequence can trace the inheritance of molecules from one stage to the next \citep{Nomura2022}. 
Within the Solar System, isotopic fractionation can trace the origins and long-term evolution of an object’s volatile inventory \citep{Cordiner2018,deKleer2024}
and place our Solar System in the context of the disk population.
ALMA has pushed the study of fractionation chemistry to rarer molecules and smaller scales than was previously possible, but so far for small and likely un-representative source samples. These measurements will be facilitated by ngVLA's unprecedented sensitivity, enabling (i) detections of rare isotopologues of large ($\gtrsim$8 atom) molecules \citep{Jorgensen2016,Burkhardt2018}, (ii) detections of low-abundance minor isotopologues or multiply substituted isotopologues \citep{Calcutt2018,Drozdovskaya2022}, and (iii) spatially resolved fractionation patterns in small molecule tracers \citep{Hily-Blant2019}.

\subsection{Telescope Requirements}

To meet the spectroscopic requirements to detect prebiotic molecules and complex organic molecules, the design of the ngVLA will need to reach rms levels of 30 uJy/beam\,km/s for frequencies between 16 GHz and 50 GHz with spectral resolution of 0.1 km/s. Angular resolution on the order of 50 mas is needed at 50 GHz and the largest angular scales expected range from $2\arcsec$ to $10\arcsec$. The unique combination of sensitivity and resolution of the ngVLA will enable better depth than any pre-existing surveys for prebiotic molecules in the Galaxy. The deepest current surveys for prebiotic molecules are being done with the GBT and the Yebes telescopes, but their spatial resolution limitations limit benefits from pushing any deeper in sensitivity. 

\subsection{Data Product and Processing Requirements:}

The most stringent data product and processing requirements are set by broadband unbiased spectral line surveys. These require the collection and imaging of the maximum available bandwidth at full spectral resolution. The most common reprocessing step for users is likely to be continuum subtraction, which is extremely challenging in line-density spectra for which few or no line-free channels are available over an entire source. Users of these surveys will wish to have the entire observed data cubes available at full spectral resolution. Self-calibration will be essential.

More targeted studies, where the target spectral lines are well-known, can work with substantially less data volume, especially if the spectral setups are sufficiently flexible. For example, when blind surveys are not required, or a target source is not line-dense, observing an entire 4 GHz chunk of spectrum at maximum spectral resolution would result in the collection of a monumental number of line-free channels. A sufficiently advanced architecture to allow dozens of individual spectral windows to be placed around target spectral lines will cut down on data volume and processing requirements substantially.  

\bigskip\bigskip\bigskip


\section{3. Key Science Goal: Charting the Assembly, Structure, and Evolution of Galaxies\\ from the First Billion Years to the Present \\}  



\subsection{Scientific rationale} 
Our modern view of galaxies is not as ``Island Universes'' but rather as ``Cosmic Ecosystems''. Galaxies are shaped by accretion (inflows), ejection (outflows), and consumption of gas. Accretion is driven on large scales by hierarchical growth of gravitationally bound structures (dark matter halos), and on smaller scales by transport of gas within galaxy disks and into galactic nuclei, driven by viscosity and dynamical processes such as mergers and internal instabilities. Molecular clouds collapse within the interstellar medium (ISM) and fragment to form stars. The resulting massive stars and supernovae deposit energy, momentum, and heavy elements into their surroundings, creating large-scale outflows that can eject gas from galaxies and even from their halos. Some gas manages to accrete all the way into galactic nuclei, fueling black holes that also inject vast quantities of energy and momentum via winds and jets. Ejected gas and metals may flow back in to galaxies, contributing to future generations of accretion. 
Although there is broad general agreement on this qualitative picture, despite rapid progress in recent years both observationally and theoretically, there are still key open questions. These include 1) How is the accretion rate of pristine gas from the IGM into the CGM related to the growth rate of dark matter halos? 2) How rapidly can gas cool and accrete from the CGM into the ISM, and what is the role of the interaction of multiphase gas with very different densities and temperatures? 3) How is gas transported within galactic disks and into the nucleus? 4) What determines the efficiency of converting dense gas into stars and how does this depend on environment? 5) What are the mass and energy loadings of stellar and AGN driven winds and jets, and how are the conditions on small scales related to outflow rates on galactic and halo scales? 6) How far do outflows travel, and what is the timescale on which ejected gas falls back in? 

\begin{figure}
\centering
\includegraphics[width=0.9\textwidth]{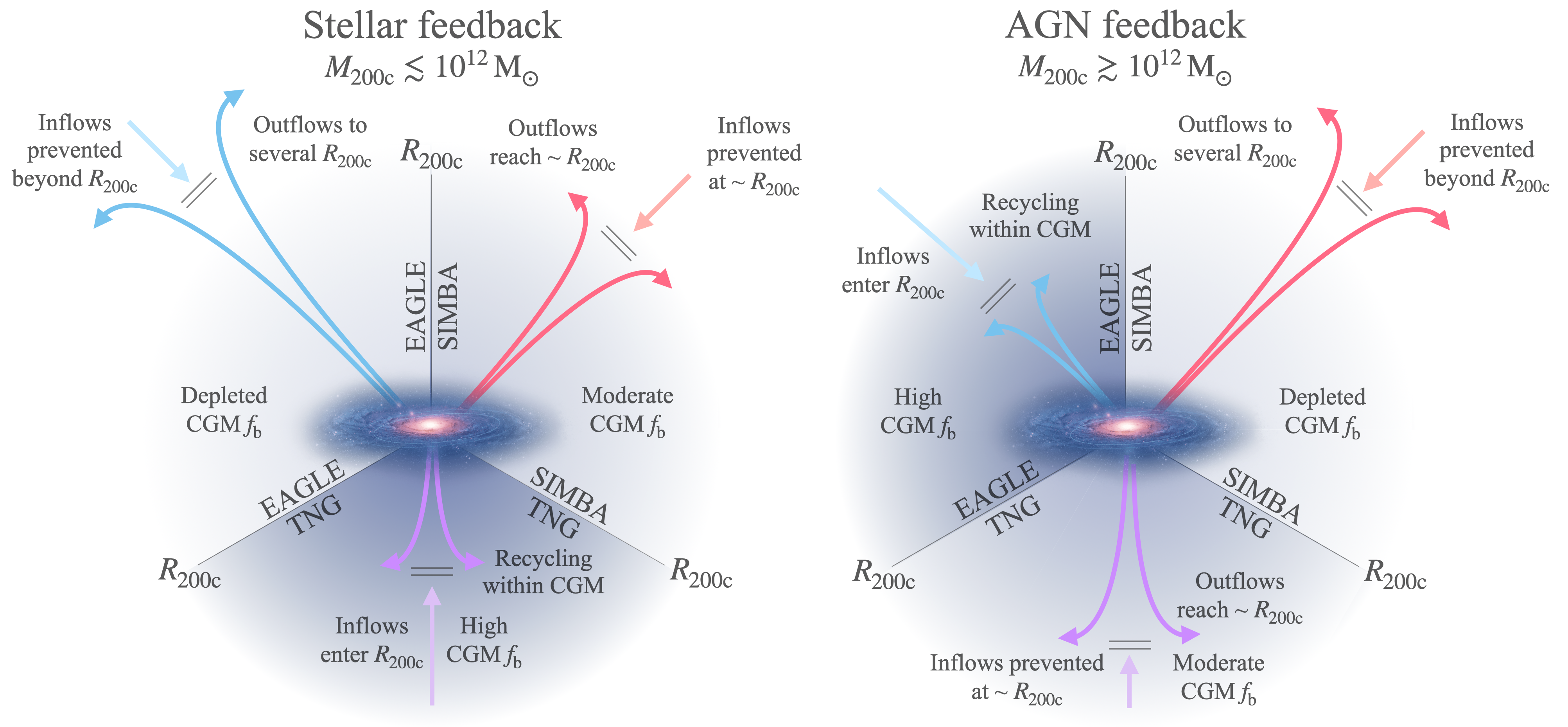}
\caption{Schematic illustration of the inflows and outflows of gas in three state-of-the-art cosmological hydrodynamic simulations. All of the simulations make similar predictions for the stellar mass function of galaxies at $z=0$, but this is accomplished via dramatically different baryon cycles. Reproduced from \citet{Wright:2024}. }
\label{fig:baryoncycle}
\end{figure}

Due to limitations of dynamic range, all theoretical simulations of galaxy evolution in a cosmological context must currently treat critical processes in the baryon cycle (e.g. star formation, stellar and AGN feedback) using ad hoc phenomenological "sub-grid" recipes \citep{Somerville_Dave:2015}. These recipes incorporate tunable parameters that are typically calibrated to reproduce global observables such as the local galaxy stellar mass function. However,  simulations that adopt different sub-grid recipes that are intended to represent the same physical processes make dramatically different predictions for the baryon cycle, as illustrated in Fig.~\ref{fig:baryoncycle} \citep{Wright:2024}. For example, some simulations adopt high mass loadings for stellar driven winds, but most of the ejected wind material does not escape the halo, and it rains back in to the galaxy on a short timescale (bottom left; purple lines). Other simulations assume energy loaded winds which predict lower mass outflow rates, but accomplish similar global star formation efficiencies by ejecting material to larger distances and by ``preventative feedback'' -- the coupled energy slows down cooling and accretion (top left; blue lines). Measurements with ngVLA of the gas content of galaxies as well as outflow properties will discriminate between these different theoretical pictures. 

The unique sensitivity, spatial resolution, and frequency coverage of the ngVLA will provide breakthrough measurements that will directly constrain these key processes, from  unprecedented studies of nearby galaxies (including our own), to the detailed characterization of galaxies at the peak of cosmic star formation (`cosmic noon'), all the way to the earliest galaxies emerging from the end of cosmic reionization (`cosmic dawn').  Indeed, the ngVLA will be able to undertake large surveys to address systematically the molecular gas content and properties in high-z galaxies using the red-shifted CO(1--0) and CO(2--1) transitions. The access to the lower level CO transitions removes the uncertainties related to the unknown excitation state of the gas, which introduce rapidly increasing biases for higher transitions (e.g., \citealt{Riechers2011, Riechers2019}).

\begin{figure}
\centering
\includegraphics[width=1\textwidth]{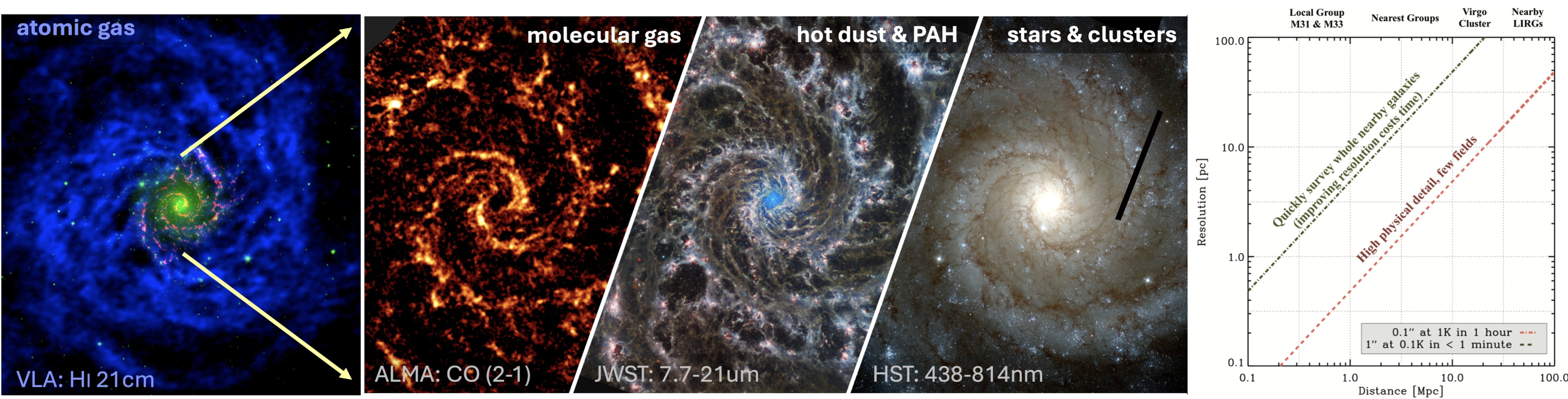}
\caption{In the nearby universe, {\bf the ngVLA will provide the highest-fidelity imaging of key gas and star formation tracers in galaxies}, from the extended atomic hydrogen reservoirs (\hi, as traced through the 21cm line, left), to the molecular gas phase, as e.g.\ traced by carbon monoxide (CO), and the stellar and dust emission, matching the ALMA/JWST/HST imaging from, e.g., the PHANGS survey \citep{Leroy21, Lee23}. Importantly, these will include matched--resolution high-density gas tracers (such as HCO$^+$ and HCN) that need the collecting area and resolution provided by the ngVLA. The right panel highlights how the ngVLA will revolutionize the studies of molecular cloud demographics in a large range of galactic environments \citep{Leroy18}.
}
\label{fig:PHANGS}
\end{figure}

\subsection{Our Cosmic Neighborhood}
Molecular clouds form via gravitational instability and cooling, and can be destroyed by dynamical processes as well as by feedback from stellar winds, photoionization, and radiation pressure. In the local universe, star formation occurs within these clouds with an efficiency of only a few percent per free-fall time, due to the feedback effects just mentioned, mediated by turbulence \citep{McKee07}. The relative importance of these processes in different environments and at different cosmic epochs is one of the key open questions in galaxy evolution. The nearby universe provides the ideal laboratory to study these mechanisms in great detail \citep{Schinnerer_Leroy:2024}. 

The ngVLA will provide the combination of surface brightness sensitivity and resolution necessary to understand the physical makeup of galactic gas reservoirs, ranging from the circum-galactic medium and the outer disks, to the detailed characterization of collapsed cores of molecular clouds, traced by CO and dense gas tracers (e.g. Fig.~\ref{fig:PHANGS}).  

The ngVLA will survey the structure of the cold, star-forming interstellar medium at the parsec-resolution of star-forming regions out to the Virgo cluster. It will image not only CO but also a host of other molecular tracers with transitions that are $1-1.5$ dex weaker, providing a range of cold interstellar medium diagnostics and mapping the motion, distribution, and physical and chemical state of the gas. The changing astrochemestry in different galactic environments also links to the core science goals of KSG\,2 ([LINK HERE]).

The ngVLA makes large systematic studies feasible for the first time, taking advantage of chemical footprints that are beyond the limits of current capabilities to yield key insights on the processes that shape galaxies. Simultaneously, imaging of free-free and synchrotron continuum emission provides the full context of star formation and accretion activity. The deep imaging of the atomic gas in the outskirts of galaxies will distinguish between settled extended \hi\ disks, in-- and out--flowing gas, merger remnants, and  tidal features. Figure~\ref{fig:PHANGS} summarizes the large range of physical properties (atomic and molecular gas, star formation) that the ngVLA will map out in detail for a vast range of galaxy environments.

\subsection{Charting the Gas Reservoirs and Kinematics of Galaxies through Cosmic Time}

\begin{figure}
\centering
\includegraphics[width=0.85\linewidth]{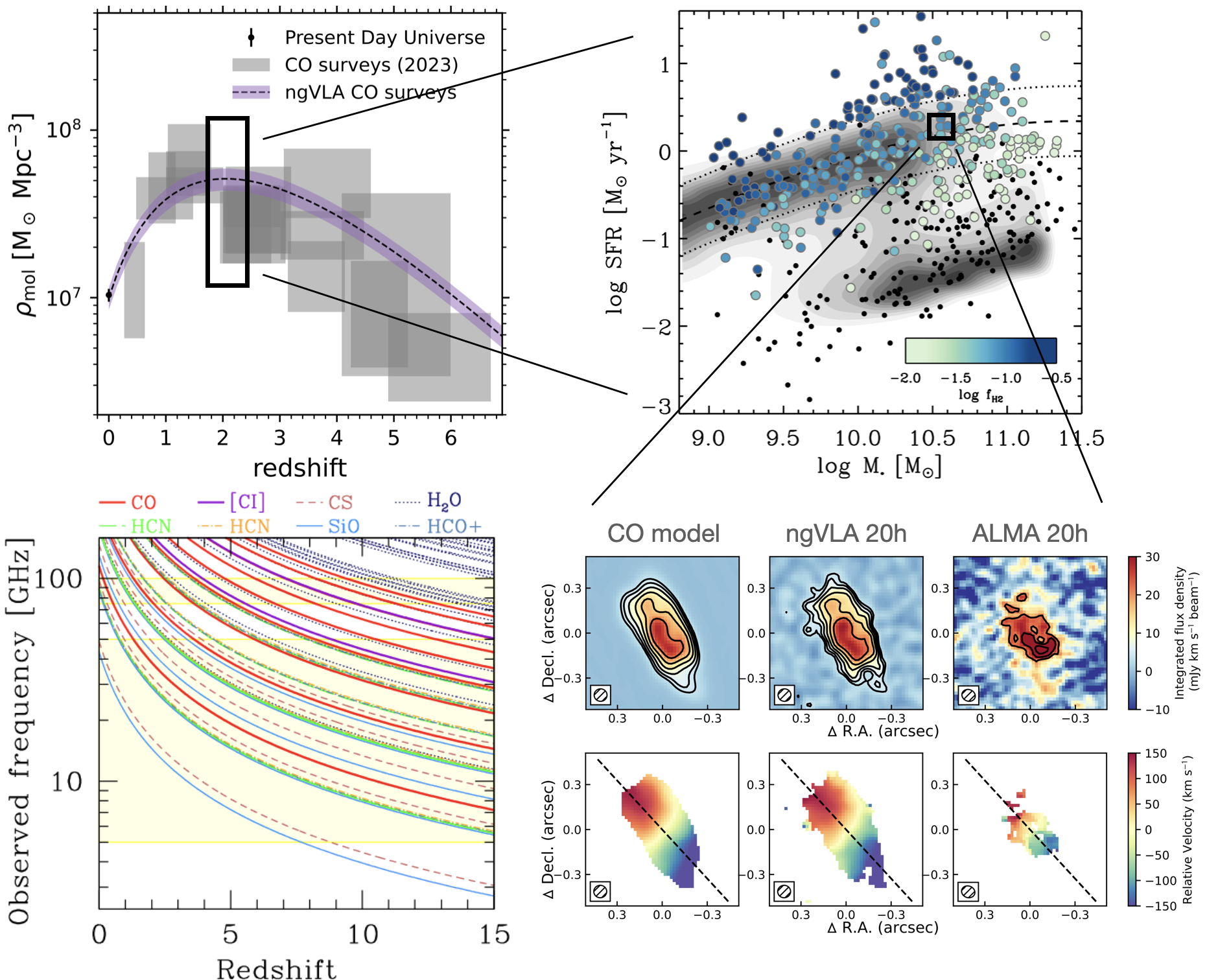}
\caption{
{\em Top left:} The ngVLA can observe the low--J CO lines out to high redshift, thus constraining the evolution of gaseous reservoirs from the epoch of reionization (`cosmic dawn') to the peak of the cosmic star formation history (`cosmic noon', adopted from \citealt{Boogaard2023}). The order-of-magnitude improvement in the number of CO detections  will enable studies of large samples of galaxies in individual redshift bins, to constrain key parameters such as gas mass fractions, that can currently only be obtained for large samples in the local universe \citep{Saintonge2017} ({\em Top right}). {\em Bottom right:} In addition to integrated measurements, the ngVLA will be able to image the cold molecular gas reservoirs in high-redshift galaxies down to sub-kpc resolution. The panel to the left shows the model distribution of CO(2--1) emission in a main sequence galaxy at z=1.7. The middle (right) panel shows a 20h ngVLA (ALMA) simulation \citep{Carilli2022}. The vastly superior collecting area of the ngVLA traces the gas much further out compared to ALMA, thus allowing for kinematic dark matter studies of large samples of high--redshift galaxies. In addition, low--J CO imaging with the ngVLA will be possible to significantly higher redshifts (where the lines shifts out of the ALMA bands). Spatially resolving the galaxy is the key to understanding the dynamical mechanisms driving star formation in galaxies during the peak cosmic star formation epoch \citep{Carilli2022}. {\em Bottom left:} Redshifted frequencies of key dense gas tracers, many of which will be accessible with a single ngVLA tuning. In particular, the 1-0 and 2-1 transitions of HNC, HCO+, HCN and SiO lines are spaced very closely in frequency. These lines are too faint for ALMA for high--redshift main--sequence galaxies, and their low--J transitions are not accessible by ALMA's frequency bands (adopted from \citealt{Decarli2018b}).}
\label{fig:ASPECS}
\end{figure}

At high redshift, the ngVLA will be able to undertake unprecedented deep, wide area observations, providing critical missing data for large multiwavelength community legacy surveys selected in the optical/IR (such as CANDELS; \citealt{Koekemoer2011}). Following the `molecular deep field' approach successfully demonstrated with ALMA \citep{Decarli2019, Decarli2020}, this will systematically address the molecular gas content and properties in and around high-z galaxies using the redshifted low--J CO transitions at throughout the last 13 Gyr of the Universe (0$<$z$<$7), down to an order of magnitude lower gas masses than currently possible \citep[e.g.,][]{Pavesi2019}. These surveys will thus expose the evolution of gaseous reservoirs from the earliest epochs to the peak of the cosmic history of star formation (Fig.~\ref{fig:ASPECS}, left) at a precision that is more than an order of magnitude better than current surveys \citep{Boogaard2023}. Such measurements then allow us to go beyond measuring quantities averaged over cosmic time and space, to statistically investigate key galaxy properties such as gas depletion times and stellar mass--to--gas ratios in specific mass and redshift bins, at a level of detail currently only achievable in studies of local galaxies \citep[e.g.][]{Saintonge2017}. Statistics of molecular detections over a broader range of galaxy masses, together with vastly increased sensitivity to continuum emission arising from synchrotron, free-free, and cold dust at high-$z$, will provide unique insights into the evolution of galaxies through cosmic time. Moreover, with the power to carry out in-depth, high-resolution imaging studies of distant galaxies, the ngVLA will make it possible to image routinely and systematically the sub-kiloparsec scale distribution and kinematic structure of molecular gas in main-sequence, dwarf, early--type, and starburst galaxies \citep{Carilli2022}, an extremely challenging task for the present generation of telescopes (Fig.~\ref{fig:ASPECS}, right; cf.\ \citealt{Kaasinen2020,Rizzo2023})\footnote{For reference, 1\,kpc corresponds to $\sim$0\farcs1--0\farcs2 at redshifts $z\sim1-10$}. Tracing the galaxy kinematics in low--excitation gas to large radii will also constrain the total dark matter content that is currently ill constrained at high redshift.
With the inclusion of the long baselines, the ngVLA will also provide the necessary sensitivity and angular resolution to infer the nature of dark matter with unprecedented precision (e.g. \citealt{Spingola18, Powell23, Vegetti23}), and spatially resolve the faintest (hence more common) galaxies that are gravitationally lensed (e.g., \citealt{Spingola20}).
The ngVLA will thus provide a key component of the combined high-resolution multi-wavelength studies that are being  undertaken JWST, ALMA, and the next generation of NASA missions, as well as the future 30\,m--class optical telescopes to chart the process of galaxy assembly throughout cosmic history. 

\subsection{Tracing the co-evolution of galaxies and supermassive black holes through Cosmic Time}

The ngVLA will also provide a unique view of the growth of accreting supermassive black holes (SMBHs) through cosmic times, through a variety of observations, ranging from the radio continuum, through outflow signatures of active galactic nuclei (AGN) to kinematic measurements of SMBH masses through maser/mega-maser studies (traced e.g.\ by  water and OH). Accretion events on SMBH will result in dynamic processes that are the core of KSG\,5 (REF). Multi-scale studies of radio jets will further constrain the role of AGN feedback in various environments.

\subsection{Telescope requirements}

An order-of-magnitude improvement in effective collecting area over the VLA at 1\,cm, and over ALMA at 3\,mm is critical to probe down to faint galaxy populations in the early universe, and enable high-resolution imaging of wide-spread, low surface brightness emission in normal galaxies. A densely packed “core” of antennas is essential to obtain the brightness sensitivity to detect molecular gas on large scales in the inter-stellar and circum-galactic medium, while also retaining the necessary point-source sensitivity and not over-resolve faint galaxies. A large instantaneous bandwidth is key to carry out efficiently blind surveys of large cosmic volumes in a single observation. At the same time, the large bandwidth will provide routine access to molecular species different than CO, such as HCN, HCO$^+$, or N$_2$H$^+$. These species can then be studied for large samples across a wide range in galaxy luminosities and masses through stacking of CO-detected galaxies. Access to transitions of formaldehyde (5 GHz and 14 GHz), ammonia ($23-27$ GHz), methanol (particularly the 36 GHz masers), deuterated molecules ($\sim70$ GHz), and a host of dense gas tracers ($\sim 90$ GHz) besides CO (115 GHz) and \hi\ (1.4 GHz) provides key probes in the nearby universe.  With full continuum capabilities the ngVLA will also be able to obtain simultaneously measurements of the star formation rate from free-free and radio recombination line emission. Accurate recovery of flux for extended objects will require placing some of the collecting area on very short baselines. Neither ALMA nor the SKA Phase~1 have the power to carry out these observations. The SKA Phase~1 will not have the frequency coverage necessary to do the vast majority of this science. Although ALMA can tackle some of these observations on a few objects, it lacks the collecting area to observe large samples, and cannot trace the critical low-J lines of CO and other molecular traces at the highest redshift.

\subsection{Data Product and Processing Requirements:}

The anticipated high level data products that the ngVLA project will deliver will be sufficient to address the key goals addressed here. For additional processing, having access to the fully calibrated visibilities through a user-facing archive would be hugely advantageous, as different science cases (e.g. bright vs.\ faint lines) will require different velocity binning, and/or tapering in the uv plane. Likewise, preserving the native channel spacing will be important for bright lines in nearby universe targets. Given the large number of relevant spectral lines, the ability to place dozens of individual spectral windows at target spectral lines would be important. Special tools will likely be needed to combine the full array with ngVLA's Short Baseline Array (SBA) to account for low surface brightness emission . Ultimately, including tools to combine ngVLA data with observations obtained at single dish telescopes (for zero spacing corrections) would be an asset. For the nearby galaxies, extra processing requirements will be needed for large--area mosaics of the skyand for creating large image cubes. Stacking routines (to stack emission from faint lines based on bright ones) could also benefit other key science projects. For continuum imaging, having access to self calibration routines would be strongly desired.

\bigskip\bigskip\bigskip


\section{4. Key Science Goal: Science at the Extremes: Pulsars as Laboratories for Fundamental Physics \\} 


\subsection{Scientific Rationale}\label{sec:tdcp.gcpsr.science}

New insights into fundamental physics are made by probing the extremes of physical systems to search for small deviations from established theories. Neutron stars and pulsars have long been identified as natural laboratories for testing physics at its extremes, especially theories of gravity and dense matter. The ngVLA will deliver next-generation insight into general relativity and the dense matter equation of state through high-sensitivity, high-frequency, and high-precision observations of pulsars, especially but not limited to those in the Galactic Center \citep{Bower2018ASPC..517..793B}. It will also advance multi-messenger astrophysics primarily through its inclusion in pulsar timing arrays for low-frequency gravitational waves \cite{Chatterjee2018ASPC..517..751C}, but also through very-long-baseline interferometry (VLBI)
to detect cosmological gravitational waves via extragalactic astrometry \citep{Darling2018ASPC..517..813D}, and potentially in tandem with space-based VLBI to directly image gravitational-wave-emitting supermassive black hole binary systems \citep[e.g., as described in][]{Carilli2023arXiv230913149C}.

Although a large population of Galactic Center pulsars is expected to exist \citep[e.g.,][]{Wharton2012ApJ...753..108W}, thus far only seven pulsars in the central half-degree of the Galaxy have been discovered.
The dearth of pulsar discoveries in this region is likely due to several factors that make these detections particularly difficult.
Pulsars are generally faint: pulsars at distances comparable to or greater than the distance to the Galactic Center represent only 20\% of the current census \citep[716/3534, including globular cluster and LMC pulsars][]{Manchester2005AJ....129.1993M}\footnote{\url{https://www.atnf.csiro.au/research/pulsar/psrcat/}}.
Nearly 75\% of these distant pulsars have been discovered in the past 15-20 years, while pulsar searches have been conducted for over a half century;
the steady increase in the discovery of more distant pulsars results from a combination of larger and more sensitive telescopes, larger bandwidth systems, and improved detection algorithms.
Additionally, the intense Galactic emission toward the inner Galaxy increases the system temperature substantially at lower frequencies, leading to searches generally being less sensitive toward the inner Galaxy.  Finally, enhanced radio-wave scattering toward the inner Galaxy further decreases the effective sensitivity of searches due to increased pulse broadening from both dispersive smearing and higher scattering timescales (longer scattering tails), although the extent and distribution of an enhanced-scattering interstellar medium is not yet known (its properties can be better probed as more pulsars are found there).

In this memo, we focus on the key science that will be accomplished with ngVLA observations of pulsars, primarily in the Galactic Center. We broadly define the Galactic Center pulsar population to be that which falls within the Central Molecular Zone with diameter $\sim$\,250 pc. A very high-reward science goal that requires the ngVLA's unmatched sensitivity at radio frequencies $\sim$\,10--15\,GHz (higher than typical pulsar observations to date) is that of finding and timing one or more pulsars orbiting the Milky Way's central supermassive black hole, Sgr A*, in order to place very tight constraints on the black hole mass and on possible deviations from general relativity. We discuss this and other key science topics in more detail below.

\subsection{A Fundamental Test of Gravity with a Pulsar Orbiting Sgr~A*}\label{sec:gravity}

That the radio source Sgr~A* represents a supermassive black hole (SMBH) at the center of the Milky Way Galaxy has been demonstrated irrefutably, by a combination of its small size \citep{1993Sci...262.1414B}, radio wavelength astrometry \citep{1999ApJ...524..816R}, by tracking stars orbiting within $\lesssim 1000$\,au of its location \citep{Ghez.1998ApJ...509..678G, Do.2019Sci...365..664D}, and, most recently, with the Event Horizon Telescope (EHT) image of Sgr~A* \citep{EHT.2022ApJ...930L..12E}.  Notably, the earliest evidence in favor of Sgr~A* being the (electromagnetic) source associated with the Galaxy's SMBH resulted from observations with the Very Long Baseline Array (VLBA).

\begin{figure}
    \centering
    \includegraphics[width=0.6\textwidth]{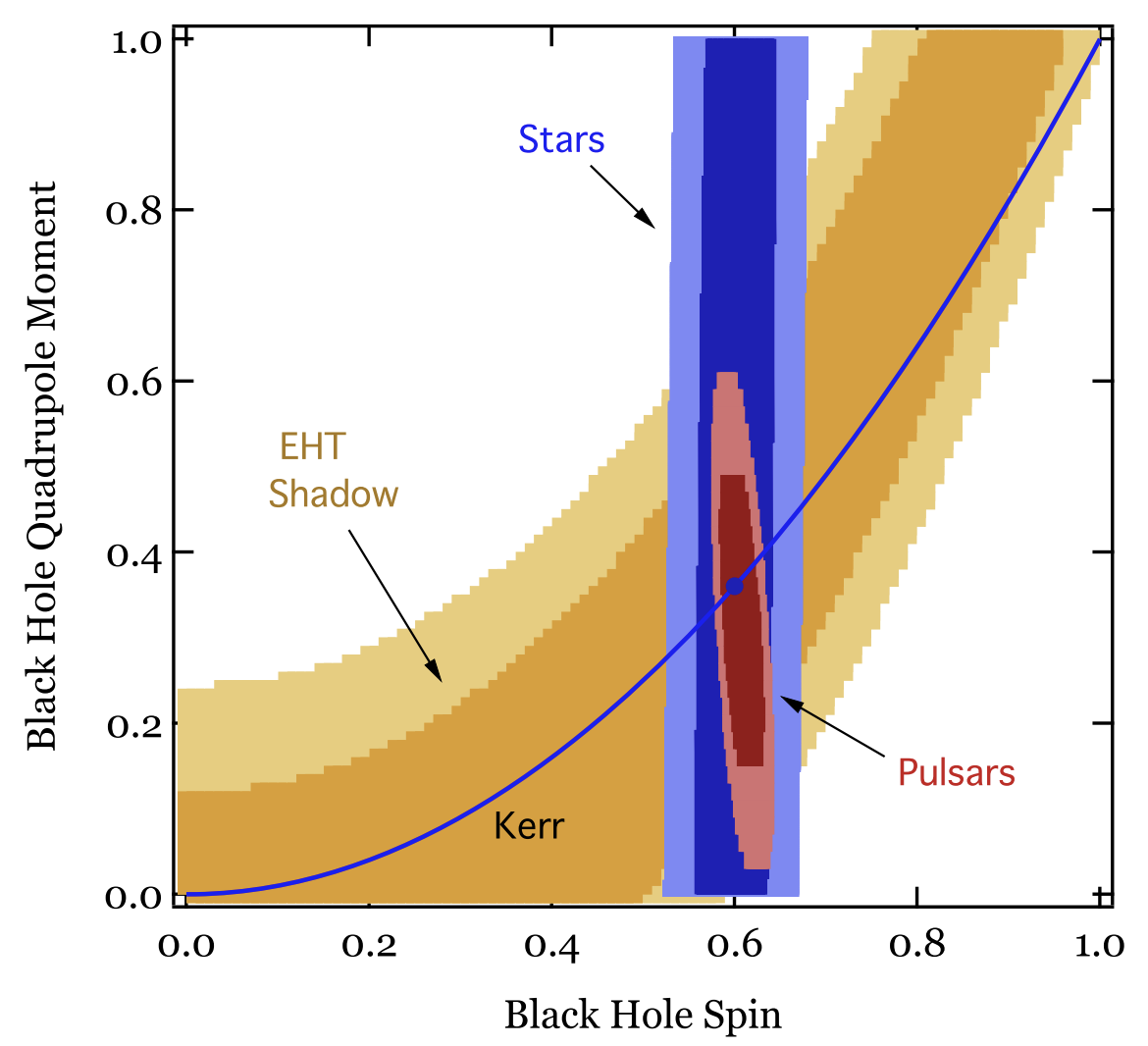}
    \caption{The quadrupole moment $q$ and spin $\chi$ of Sgr A*, where the solid curve is the $q-\chi$ relation predicted by the Kerr metric. Just one orbiting pulsar (model posterior points shown in red) would provide far stronger constraints than NIR observations of multiple stars (blue), or modelling of sub-mm VLBI images data (yellow; based on predictions made prior to EHT observations). Figure taken from \citet{Psaltis2016ApJ...818..121P}.}
    \label{fig:psaltis}
\end{figure}

Just as there are stars orbiting Sgr~A*,
the putative location of the \hbox{SMBH},
it is likely that one or more pulsars are also orbiting the black hole, given the large number of neutron stars and pulsars expected in the Galactic Center.
Such pulsars would represent clocks moving in the spacetime potential of the SMBH. Due to the precision achieved with pulsar timing, observations of Galactic Center pulsars, especially those orbiting Sgr~A*, would enable qualitatively new tests of theories of gravity (Figure~\ref{fig:psaltis}). In particular, timing a pulsar orbiting Sgr A* would yield an incredibly small black hole mass uncertainty of $\sim$\,1\,$M_{\odot}$, would test the cosmic censorship conjecture
to a precision of about 0.1\%, and would test the no-hair theorem to a precision of $\sim$\,1\% \citep{Liu2012ApJ...747....1L, Bower2018ASPC..517..793B}. Notably, given 
the strong space-time potential near Sgr~A*, it is possible for even canonical pulsars (with spin periods $\sim 1$~s) to provide such highly-constraining measurements.

Additionally---and relevant to pulsars orbiting Sgr A* as well as to other Galactic Center pulsars---high precision astrometry with the ngVLA's long baselines can be used to measure pulsar proper motions and parallaxes to higher precision than is possible with pulsar timing, especially in the case of a canonical pulsar in the immediate vicinity of Sgr A*.  Providing these astrometric parameters as priors to the pulsar timing model can reduce parameter covariances and improve the overall timing precision: past efforts with the VLBA have yielded improvements to GR tests (e.g., Kramer et al., 2021, Ding et al., 2023) but the enhanced sensitivity of the ngVLA will open up many more systems to VLBI astrometry.

\subsection{The \textit{Fermi} GeV Excess: Dark matter annihilation or a population of millisecond pulsars in the Galactic Center}

The \textit{Fermi} Large Area Telescope (LAT) revealed an excess of GeV emission toward the Galactic Center that could potentially be a signature of dark matter annihilation \citep[e.g.,][and references therein]{Ackermann2017ApJ...840...43A}.
However, as more and more millisecond pulsars (MSPs) were discovered at the positions of LAT ``unassociated'' sources (primarily point sources with no known multi-wavelength counterpart), and pulsed gamma-ray emission was found to be ubiquitous in the MSP population, an alternative explanation for the emission was proposed: that the excess gamma-rays were being produced in the magnetospheres of an as-yet-uncovered population of Galactic Center MSPs \citep[e.g.,][and references therein]{Bartels2016PhRvL.116e1102B}. While other upcoming facilities such as the SKA will have high enough sensitivity to discover a portion of this population, the ngVLA will provide the key capability of piercing through high-scattering regions, likely especially within the Central Molecular Zone, with its high-frequency coverage. If a large MSP population is found in the Galactic Center region, then the dark matter origin theory for the gamma-ray excess would be disproved. Alternatively, if even the ngVLA does not detect a sizable Galactic Center MSP population, then this would make dark matter annihilation a more tantalizing explanation for this emission.

\subsection{Pulsar-Black Hole Binary Systems and Compact Pulsar Binaries}\label{sec:binaries}

Beyond the Galactic center, the ngVLA capabilities would enable another approach to probing gravity.  First, it is widely expected that the Galaxy may contain a small number of pulsar-black hole binaries \citep{Lipunov2005MNRAS.359.1517L, OShaughnessy2008ApJ...672..479O}.  Likely even more powerful than relativistic neutron star-neutron star binaries for testing theories of gravity \citep{Liu2014MNRAS.445.3115L}, some pulsar-black hole binaries may form in the Galactic center region itself, due to three-body interactions resulting from the high stellar density in the region, in an analogous manner to how relativistic binaries are formed in globular clusters (which themselves may also host pulsar-black hole binaries).  Another formation channel, though, is through normal stellar evolution of a high-mass stellar binary.  Regardless of formation channel, if any of these binaries exist in the Galaxy, they are likely to be rare and therefore distant (for example, in the current census of $>3500$ pulsars, no pulsar-black hole binaries are known, and there are $\lesssim$\,20 known neutron star-neutron star binaries).  Consequently, any pulsar-black hole binaries in the Galactic disk could experience significant pulse broadening from dispersion and scattering, even if not as severe as that for lines of sight to the Galactic Center itself.

Moreover, a significant limitation to finding highly relativistic binaries---pulsar-black hole binaries as well as compact pulsar binaries, some of which may be useful for dense matter studies as described below---are the accelerations experienced by the pulsars.  The ngVLA imaging capabilities open up hybrid approaches to finding pulsars, conducting a search first for compact sources (potentially with steep spectra), followed by a targeted periodicity search on candidates. For example, \citet{Bhakta2017MNRAS.468.2526B} used this hybrid imaging-periodicity technique in their recent successful detection of the recycled pulsar PSR~1751$-$2737.  The enhanced sensitivity of the ngVLA at radio frequencies of~3~GHz to~30~GHz will open a new door for the discovery and study of pulsars not only in orbit around Sgr~A*, but throughout the inner Galaxy.

\subsection{The Nuclear Matter Equation of State}\label{sec:eos}

Precision timing of pulsars in binary systems can place stringent constraints on their masses.
These mass estimates themselves provide constraints on the nuclear matter equation of state (EoS) that complement ground-based collider experiments, which do not approach the nuclear densities that occur at the center of a neutron star.
Further, these mass estimates derived from pulsar timing are a crucial element of multi-messenger studies of the extremes of nuclear matter when they are combined with thermal X-ray measurements \citep[yielding tighter constraints on radius measurements derived from modeling X-ray pulse profiles observed with \textit{NICER}, e.g. most recently][]{Dittman2024arXiv240614467D} and gravitational wave measurements \citep[tidal deformability from LIGO-Virgo, e.g.][]{Shao.PhysRevX.7.041025}. The ngVLA's sensitivity, including when combined with the hybrid imaging-periodicity search technique described above, will allow the discoveries of more binary pulsars, especially compact systems (both in and out of the Galactic Center) with which relativistic effects, and therefore pulsar masses, could be measured.

\subsection{Studying the Nanohertz Gravitational Wave Sky with Pulsar Timing Arrays}\label{sec:nHz}

The ultra-low-frequency gravitational waves expected to be generated by supermassive binary black holes (or other more speculative sources such as cosmic strings) can be detected and studied through their impact on the arrival times of pulses from radio pulsars \citep[e.g.,][]{Verbiest2024ResPh..6107719V}. The first evidence indicating a stochastic background of nHz gravitational waves has recently been presented by several pulsar timing array consortia  \citep{NANOGrav15,EPTA,PPTA}, and the next decade should offer the opportunity to use this new window to study galaxy assembly across the history of the Universe.

As the study of the nHz gravitational wave sky turns from detection to characterization, including the detection of individual gravitational wave-emitting SMBH binaries \citep[if they are a significant contribution to the stochastic background, then several individual foreground binaries are expected to be detected in addition to the background; e.g.,][]{Gardiner2024ApJ...965..164G}, the ngVLA can contribute in two transformational ways: 
\begin{enumerate}
     \item By measuring the distance to a handful of the nearest, brightest, and best-timed pulsars to sub-parsec accuracy (i.e., a fraction of a wavelength of the gravitational waves of interest) via annual geometric parallax, ngVLA will enable the ``pulsar term" of the gravitational wave (i.e., its effect at the location of the pulsar) to be coherently included as a signal source when modeling discrete GW sources, rather than contributing additional unmodeled noise. Ultimately, this would lead to the use of pulsar timing arrays as true imaging interferometers of the nHz GW sky, and is {\bf only} possible due to the combination of sensitivity, field of view, frequency, and baseline length of the \hbox{ngVLA}.


     \item In terms of the pulsar timing itself, the ngVLA's sensitivity will allow many new (faint) pulsars to be added to the timing array and improve the sensitivity of those already being timed \citep{2018ApJ...861...12L}.  Notably, once the nHz GWs have been detected and are being characterised, the number of pulsars becomes more important than the precision of the best pulsars \citep{Siemens2013}.
\end{enumerate}

\begin{figure}
    \centering
    \includegraphics[clip, trim=0cm 2.0cm 0cm 2.0cm, width=0.6\textwidth]{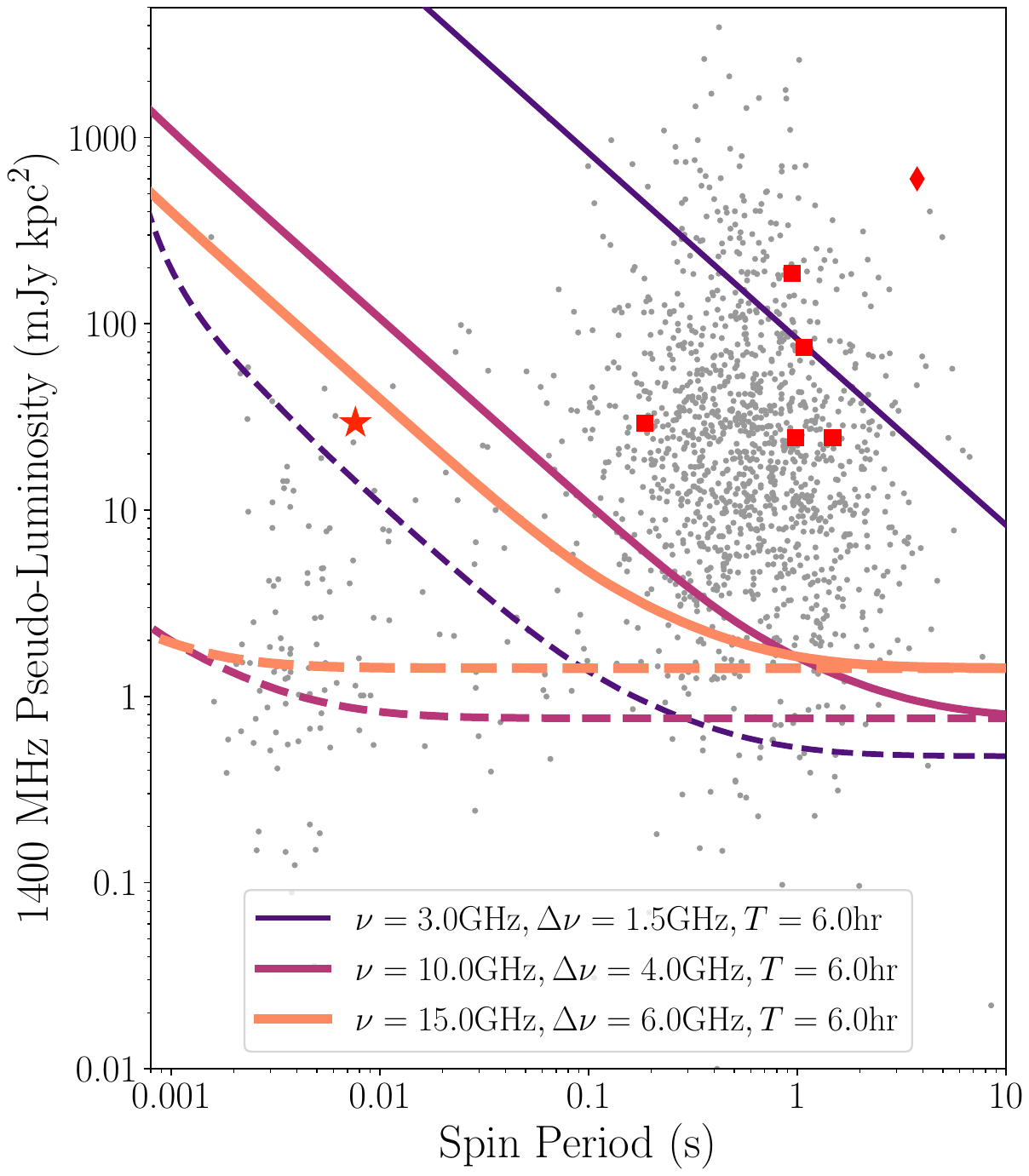}
    \caption{The sensitivity of the ngVLA to the known Galactic pulsar population when placed at the distance of the Galactic Center, under the assumption that the ngVLA will be $\sim$\,10$\times$ as sensitive as the VLA at the frequencies listed in the figure legend. Solid lines assume hyper-scattering in the Galactic Center, while dashed lines assume the scattering timescale observed in the magnetar PSR~J1745$-$2900. The red points represent the five canonical pulsars (squares), the magnetar (diamond), and the millisecond pulsar (star). Figure credit: R. Wharton.}
    \label{fig:ngvla_sensitivity}
\end{figure}

\subsection{Unveiling the Galactic Center pulsar population with the ngVLA}

There are multiple reasons to expect that the Galactic Center region contains a large number of neutron stars, ranging
from the observed population of young, hot stars, to candidate pulsar wind nebulae and X-ray binaries, to estimates of the supernova rate derived from the diffuse X-ray emission. In addition to young or canonical pulsars, a large population of MSPs may also exist in the inner Galaxy; the resulting estimates for the number of active pulsars beamed toward the Earth are as high as $\sim$\,1,000 \citep{Wharton2012ApJ...753..108W, Eatough2015aska.confE..45E}. Uncovering a substantial fraction of this population will provide insights into star formation history in the Galactic Center (via pulsar characteristic ages), the properties of the interstellar medium, the magnetic field via Faraday rotation, and the local gravitational potential by using MSPs as accelerometers \citep{Bower2018ASPC..517..793B}.

As discussed previously, the reasons that so few Galactic Center pulsars have been detected so far include pulsars' low flux density, especially at higher observing frequencies due to their steep radio spectra, and dispersion and scattering, which can broaden pulses to the point that they are undetectable. It has long been suggested that extreme scattering in the ISM of the Galactic Center is responsible for the lack of pulsars discovered there, but the true distribution of high-scattering material is not yet known. The seven Galactic Center pulsars that have been detected to date \citep{Johnston2006MNRAS.373L...6J, Deneva2009ApJ...702L.177D, Shannon2013MNRAS.435L..29S, Lower2024ApJ...967L..16L} show varying degrees of scattering: for example, the first magnetar (a highly magnetized pulsar) discovered near Sgr~A*, PSR~J1745$-$2900, shows substantial pulse broadening relative to most other pulsar lines of sight, but the scattering timescale is significantly lower than original estimations \citep{Shannon2013MNRAS.435L..29S}; and the only known Galactic Center MSP has an extremely low scattering timescale of only $\tau_{\mathrm s} \sim 0.87$\,ms \citep{Lower2024ApJ...967L..16L}. One possibility is that the Galactic Center ISM is clumpy, such that some lines of sight have significantly higher scattering timescales than others, which would explain the lower-than-expected scattering in the pulsars that have been detected (i.e. they happen to lie along low-scattering lines of sight and thus were detectable at $\lesssim$\,few GHz with existing radio telescopes).

Finding many Galactic Center pulsars (along many lines of sight) will allow us to probe the scattering characteristics of the ISM there. The ngVLA is crucial for finding these pulsars: observing at higher frequencies than are planned for the SKA can mitigate radio-wave scattering, but without an accompanying increase in sensitivity the benefits are limited due to the generally steep radio spectra of pulsars. The ngVLA's combination of a dramatic increase over present collecting area---thus providing the necessary increase in sensitivity---and the capability to observe at higher frequencies is therefore necessary for these searches. Figure~\ref{fig:ngvla_sensitivity} demonstrates the ability of the ngVLA to uncover much of the Galactic Center pulsar population.

\bigskip

It is important to acknowledge the possibilities that (a) no pulsars will be discovered orbiting Sgr~A* and/or (b) few Galactic Center pulsars will be found. In the unfortunate event that a new GR test cannot be done with a pulsar orbiting the Milky Way's central black hole, there is still much science that can be done with pulsars discovered elsewhere in the GC, as discussed above. And, in the unlikely scenario where no or very few pulsars are found in the GC, then this would raise the tantalizing suggestion that something about star formation and evolution, the GC pulsar population, or the Galaxy's dynamical past, is very different from what has been predicted thus far.

Finally, it is worth noting that the ngVLA will complement other large facilities, in particular the SKA, that will be operating in the same era. The SKA will have a large field of view and will be an excellent telescope for pulsar searching at $\sim$\,1--few GHz, while the ngVLA will provide high sensitivity at much higher frequencies, where pulsars are fainter but significantly less scattered. The ngVLA may also be more sensitive to highly-scattered, flat-spectrum magnetars. Thus the SKA and ngVLA will be complementary in the interplay between spectral and scattering properties of the pulsars they will detect.

\subsection{Telescope Requirements:}\label{sec:telescope_reqs}


The search for and study of Galactic Center pulsars is likely to drive the design of the ngVLA in three aspects: frequency range, sensitivity, and signal processing.

For the frequency range, scattering is the primary factor. While there are uncertainties in the distribution of scattering electrons in the ISM, and additionally the distribution could be inhomogeneous, in order to mitigate radio-wave scattering it is likely that pulsar studies will require a frequency range that includes the lower range anticipated for the ngVLA ($>3$\,GHz).

A significantly higher sensitivity than the VLA is important for detecting pulsars in the Galactic Center. Current searches with 100-m-class radio telescopes have found few pulsars, indicating that substantial additional sensitivity is necessary. An order-of-magnitude sensitivity would open the doors to new pulsar discoveries.

Specific signal processing requirements include beam-forming capability and pulse-binned imaging. To achieve this Key Science Goal, a pulsar survey of the Galactic Center will be needed; and the scientific reward from Galactic Center (and all other) pulsars results from precision timing. A beam-forming capability will be required for both the pulsar survey and pulsar timing. In particular, to perform the Galactic Center pulsar survey in a reasonable length of time, we require $\sim 10$ beams. Additionally, the ability to perform pulse-binned imaging is important, especially for astrometric studies. This capability is currently available on the VLA and VLBA.

\subsection{Data Product and Processing Requirements:}\label{sec:tdcp.gcpsr.reqs}

\noindent Here we specify minimum requirements for data products and processing necessary to maximize the scientific return of this Key Science Goal. Many of the requirements described here are also directly relevant to other
pulsar-related ngVLA science.

\textit{Data products:} For search-mode data, providing the user with raw data is sufficient.
For fold-mode data, in addition to providing the user with the raw data, they should also be provided with flux- and polarization-calibrated pulsar profiles in {\sc psrfits} format (which is a standard and widely-used format for pulsar data). These profiles should have the same time and frequency resolution as the original observations. While not necessary, an additional set of profiles that have also had RFI cleaning performed on them would be useful in many situations. Additionally, the correlator should be able to produce pulse-binned visibility data.

\textit{Processing:} In order to supply the above fold-mode data products, a pipeline that performs the calibration will need to be run on all of the fold-mode pulsar data. This pipeline could be modeled after
\texttt{nanopipe}, which is already used by the NANOGrav collaboration. Additionally, this science case will require processing large amounts of search-mode data in order to find pulsars orbiting Sgr A*, as well as the ability to time any pulsars that are found in order to achieve the actual science goals. Thus KSG4 also requires: adequate storage space for raw search-mode pulsar data; the ability to run pulsar searches and perform common analyses like calibration and timing on the ngVLA machines, thus requiring that standard pulsar search, calibration, timing software, and pipelines be installed and maintained there (including but not limited to {\sc PRESTO}, {\sc PSRCHIVE}, {\sc Tempo}, {\sc Tempo2}, and {\sc PINT})\footnote{Software links: \texttt{nanopipe}, \url{https://github.com/demorest/nanopipe}; {\sc PRESTO}, \url{https://github.com/scottransom/presto}; {\sc PSRCHIVE}, \url{https://psrchive.sourceforge.net/index.shtml}; {\sc Tempo}, \url{https://github.com/nanograv/tempo}; {\sc Tempo2}, \url{https://github.com/mattpitkin/tempo2}; and {\sc PINT}, \url{https://github.com/nanograv/PINT}.},
any necessary software for pulse-binned or -gated imaging, which allows precise determination of the pulsar's position and is thus especially important for astrometry, and also allows searches for off-pulse emission from eclipsing pulsars and other pulsars that may emit outside their main pulse; and any necessary software for beam-forming, which is needed for pulsar search and timing observations.

\bigskip\bigskip\bigskip
%

\section{5. Key Science Goal: Understanding the Formation and Evolution of Stellar and Supermassive Black Holes in the Era of Multi-Messenger Astronomy \\}
\label{sec:tdcp.astro}

\subsection{Scientific Rationale}\label{sec:tdcp.astro.science}

They are centrally involved in most cataclysmic events in the universe and feature prominently as sources in multi-messenger astronomy.
Black holes impact the environment in which they are found; since they exist across a spectrum of mass, from stellar mass to objects more than 10 billion times as massive as the Sun (hereafter, SMBH), they have an outsized effect on processes up to and including galaxy formation and evolution.
Studies of black holes at radio wavelengths provide crucial information not just about the compact objects themselves and details about their powerful jets, but how they formed, the type of impact they are having and will have on the galaxy in which they live.

Relativistic jets launched by stellar mass and SMBHs are sources of electromagnetic emission in the radio band, and as such represent key targets for the ngVLA. 
These ejecta can  strongly affect galaxy formation and evolution \citep[e.g.][]{Silk2014}, and are thought to produce high energy cosmic rays and high energy neutrinos \citet{Heinz2002,IceCube2018}.
Probing their emission in the radio can unveil the physics behind their black hole progenitors, and the properties of their surroundings.
The ngVLA will add top-notch capabilities in terms of sensitivity, survey speed, localization capability and frequency agility, as well as resolution for ejecta studies.
With the ability to observe across a range of frequencies from 1 GHz to 116 GHz, the ngVLA will be able to target the full gamut of explosive events occurring in a range of diverse environments, from neutron star mergers (peaking at $\sim$1.5 GHz) to mergers of SMBHs (peaking at $\nu$ $\gg$ 10 GHz).

The recent decadal survey in astronomy and astrophysics laid out three science themes which are ripe for development in the coming decade, with black holes figuring prominently in two of them.
The first and most relevant, New Messengers and New Physics, recognizes the growth and flowering of observational capabilities to study extremes of gravitation and acceleration.
The ngVLA will come on-line at the culmination of a phase of rapid growth in gravitational wave and neutrino astronomy capabilities (e.g., LIGO, Virgo, KAGRA, LIGO India, Voyager, Cosmic Explorer, Einstein Telescope, LISA, NANOGrav, EPTA, PPTA, IPTA, PINGU, IceCube Gen2, KM3NeT, etc.).
The detection of electromagnetic radiation— and specifically radio emission— from energetic, and often cataclysmic multi-messenger sources associated with black holes, can enable their precise localization, establish energetics, and provide clues on their surrounding environments.
It is the combination of multi-messenger information that will provide a complete picture of the life-cycle of massive stars, the micro-physics of their explosive deaths, and the formation and evolution of neutron stars, stellar black holes, and SMBHs.
What we know about the electromagnetic counterparts of gravitational wave and neutrino events, and their frequency evolution, indicates that the ngVLA will be a key component of the rapidly developing landscape of multi-messenger astronomy.

The second relevant science theme from the astronomy and astrophysics decadal survey, Cosmic Ecosystems, has as a key component the exploration of the presence and impact of black holes on their environment.
Radio emission will be unaffected by dust obscuration, and the ngVLA will have the angular resolution to separate Galactic sources from background objects using proper motions.
The ngVLA, with its high sensitivity and high angular resolution, will enable a census of black holes on all scales, from stellar-mass to supermassive, contributing a key voice in all of these outstanding topics.

\subsection{Black holes across the mass scale}

We now know that black holes exist on practically all mass scales, however the astrophysics of how these objects form and grow remains a mystery.
LIGO is now detecting black holes that are substantially more massive than previously known stellar mass black holes
\citep[e.g.][and references therein]{Abbott2021a}, and observing black hole-black hole mergers—although we do not know how stellar-mass black hole binaries form. 
Black holes intermediate in mass between stellar mass and SMBHs (hereafter, IMBH) have long been theorized, although active searches for them have not produced conclusive results.
While SMBHs are thought to be widespread in galaxy centers, we do not understand how their growth was seeded or how (and how often) these extreme objects coalesce.
SMBHs in binary systems present their own fascinating physics as intense low-frequency gravitational-wave emitters and potential locations of circumbinary disks, and of single and dual active nuclei in an extremely dynamic environment.

In the Milky Way Galaxy, the number of X-ray binaries (containing stellar mass black holes) is only weakly constrained to be somewhere in the range 10$^{2}$–10$^{8}$ \citep{Tetarenko2016}, based on a small sample of just 20–30 known stellar-mass black holes \citep{Corral-Santana2016}.
Detection and further study requires superb localization capabilities not affected by intervening interstellar dust.
The ngVLA will have the angular resolution to separate Galactic sources from background objects using proper motions.
An ngVLA survey of the Galaxy could detect jet-powered synchrotron emission from weakly accreting black holes, increasing the black hole sample by at least an order of magnitude
\citep{Maccarone2018}.
Detailed radio studies of these objects have long been used to study accretion physics and jet formation
\citep[e.g.][]{Fender2010}.
Key information on jet structures and energy conversion mechanisms arises from rapid multi-wavelength variability measures of time lags and correlations across a range of wavelengths, but such studies are currently strongly sensitivity-limited.
Operating the ngVLA with a large number of subarrays allows continuous source and calibrator coverage over a wide range of frequencies, crucial to making advances in understanding jet physics.

IMBHs represent the missing link between the stellar mass black holes and SMBHs. 
The existence of IMBHs with masses in the range 10$^{2}$–10$^{4}$ M$\odot$ has been a long-standing puzzle in astronomy
\citep{Greene2020}.
Theory suggests that globular clusters (GCs) of stars can host IMBHs with masses 10$^{2}$-10$^{5}$ M$\odot$
\citep[e.g.][]{Miller2002}.
The sensitivity and angular resolution of the ngVLA will be key to making a breakthrough inventory of IMBHs in hundreds of globular cluster systems out to a distance of 25 Mpc \citep{Wrobel2018}.

By enabling sensitive, high-resolution imaging of a large number of targets, the ngVLA can provide unprecedented searches for and studies of SMBHs in dual active galactic nuclei.
Direct imaging of binary systems with high-resolution radio imaging has been heralded as the most confident electromagnetic method of identifying otherwise elusive binary SMBH systems \citep[Fig. 3; see also][]{Bogdanovic2022}.
The raw sensitivity of ngVLA will enable imaging of dual active nuclei at parsec to Mpc scales, in a small fraction of the time currently needed for such studies.
With extended (8000 km) baselines, rigorous searches for binary SMBHs may be carried out.
Large samples of SMBH binaries detected with the ngVLA will allow the determination of their merger rate, and the efficiency of binary evolution from kpc to sub-pc scales (the latter may be measured via flux or jet variability, or via direct imaging of two cores for
z $\lesssim 0.2$ systems, assuming 8000 km baselines are available).
Overall, the ngVLA will survey the population of binary SMBHs at a range of separations, identifying a large sample of systems with which to study the evolution of, and identify multi-wavelength emission associated with, dual and binary SMBHs.

Indeed, pulsar timing arrays operating in the nanohertz to microhertz gravitational wave-band, are on-track to begin detecting gravitational-wave sources this decade \citep{Mingarelli2017}, and the ngVLA will be the leading tool for radio follow-up of binary SMBH candidates via flux monitoring, and high-resolution imaging.
These measurements can directly constrain binary SMBH contributions to the stochastic nanohertz gravitational wave background.
Recently, \citet{Bansal2017}  made the first measurements of the orbit of a SMBH binary with the VLBA (Figure 5).
With greater sensitivity, there will be a far greater chance to identify and measure the proper motions of double AGN in much tighter orbits, where precision tests of General Relativity could be made; potentially, a multi-messenger source (trackable by the ngVLA, but also detectable as a gravitational-wave source by pulsar timing arrays) may be identified by the ngVLA via time-resolved tracking of a double core or the motion of helical jets.

\subsection{From detection to formation channels}
While we have a rough understanding of how to make single stellar mass black holes, recent LIGO results reveal the extent of our ignorance regarding how binary black holes are formed.
Simply measuring the size of the black hole population in a given mass range would have profound implications for key parameters impacting binary black hole formation, such as common-envelope evolution and the strength of dynamical “kicks” delivered to black holes at birth (caused by asymmetries in the parent core-collapse supernovae).
These parameters are key inputs into one of the core problems that has already developed in gravitational wave astronomy — whether double black holes form through normal binary stellar evolution \citep[e.g.][]{Belczynski2016}, more exotic channels of binary evolution \citep[e.g.][]{deMink2016}, dynamical formation in globular clusters \citep{Banerjee2010,Rodriguez2015}, dynamical formation in accretion disks \citep{Bartos2017}, or some combination of the above.
Binary evolution predicts a variety of scenarios in which compact objects accrete and are detectable as radio sources.
The black hole natal kicks, along with the distribution of black hole masses and spins, also provide one of the few direct means to test the mechanisms by which black holes form.
The ngVLA will directly measure black hole natal kicks through determination of proper motion and parallax.

\begin{figure}
    \centering
    \includegraphics[scale=0.5]{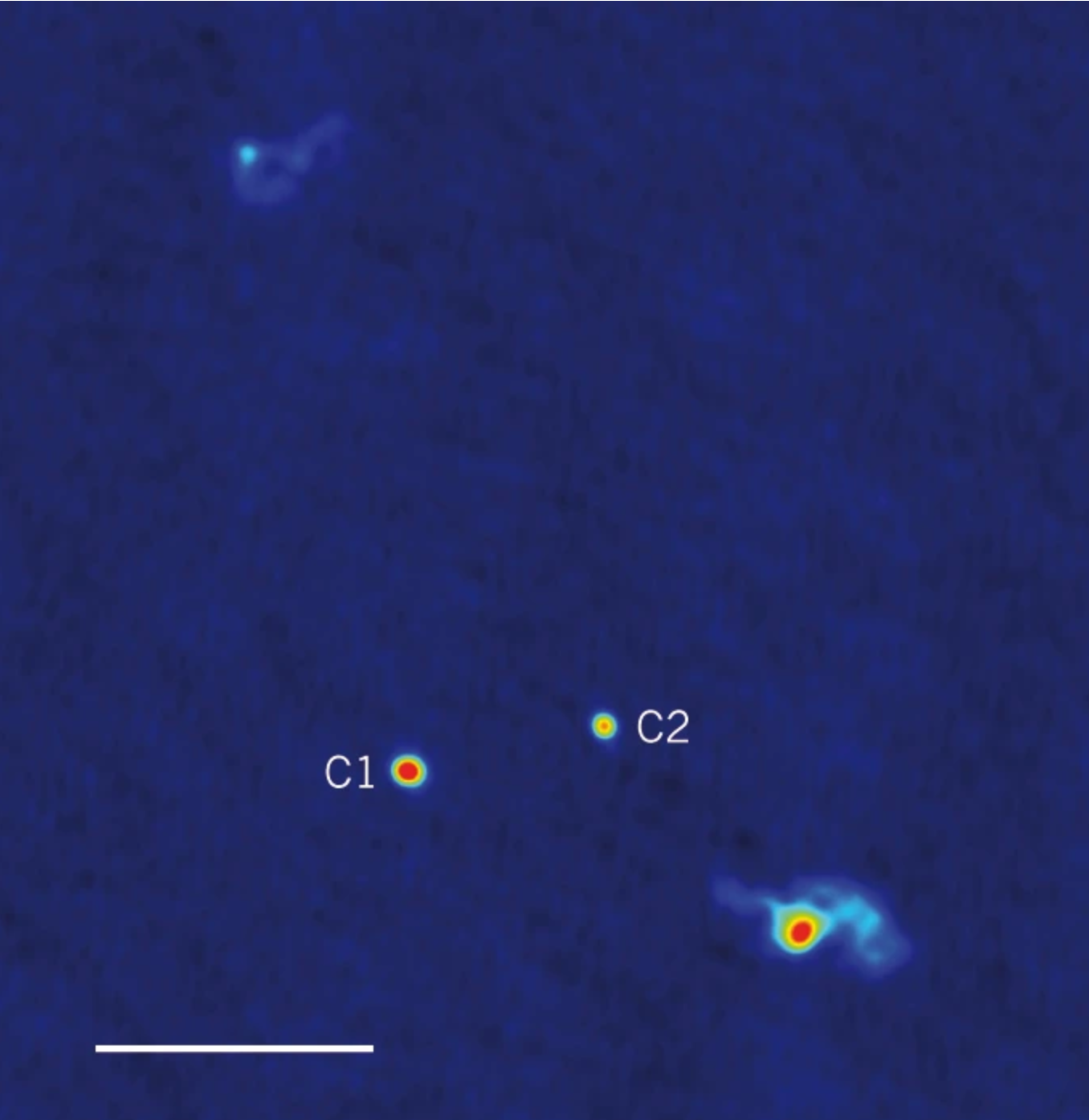}
    \caption{The ngVLA will be an excellent tool for hunting black holes, including binary supermassive black holes. Here we show a binary system of SMBHs at z=0.06. The black holes have a projected separation of 7 pc (the white scale bar denotes 10 pc) with an orbital period of 30,000 yr. Jet emission extends from the black hole C2. Past searches to identify such targets are thought to be limited by sensitivity and/or resolution \citep[e.g.][]{Burke-Spolaor2011,Tremblay2016,Breiding2021}. The ngVLA, with its deep high resolution imaging capabilities, will enable discovery of many more such systems, with intimate synergies to LISA and Pulsar Timing Arrays. Image from \citet{Taylor2014}}
\end{figure}

There is currently a mass gap between well-understood and -studied stellar-mass black holes and well-studied SMBHs \citep[e.g.][]{Tetarenko2016}. IMBHs are neither well-studied nor well-understood.
Confirming the existence of intermediate mass black holes would have immediate relevance – finding them in globular clusters would validate a formation channel for seed black holes in the early universe \citep[e.g.][]{Volonteri2010}.
Additionally, their presence and properties enable tests of whether scaling relations between stellar systems and their central black holes extend into these poorly-explored mass regimes \citep[e.g.][]{Graham2015}, and inform event predictions for gravitational wave facilities.
With its high sensitivity, high spatial resolution, and wide field of view, the ngVLA will search for the radio synchrotron signatures of slow accretion onto putative IMBHs in the hundreds of globular clusters that surround each massive galaxy in the local Universe \citep[e.g.][]{Wrobel2021}.

The ngVLA will also contribute to studies of SMBH binary formation and evolution in critical ways.
Leading models by which black holes grow to masses of order 10$^{6}$–10$^{10}$ M$_{\odot}$. include the merger of less massive black hole “seeds” (i.e., remnants of Population III stars) and direct collapse of more massive black holes in early dark matter halos \citep{Volonteri2003}, followed by continued major (similar-mass-ratio) and minor (large-mass-ratio) mergers of host galaxies, and ongoing accretion of ambient stars and gas.
Black-hole seed models can be best tested by measuring the occupation fraction of SMBHs in nearby, low-luminosity galaxies \citep[e.g.][]{Mezcua2017}.
Deep, high-resolution imaging of nearby galaxies with the ngVLA will provide proper motion distinction between nearby, low-luminosity active galactic nuclei and background sources on scales of about $\sim$10     $\mu$as per year
\citep{Plotkin2018}.

\subsection{Black holes as time-domain and multi-messenger sources}

The phase space of relativistic explosions that are fast-evolving and radio dim \citep{Soderberg2006} is currently largely unexplored, and ngVLA’s sensitivity would enable orders of magnitude improvement in detection of these types of transients.
Jets launched by newly-born stellar mass black holes, either as single objects formed in core-collapse supernovae (SNe) or long gamma-ray bursts (GRBs), or in binary systems (black hole - neutron star and black hole - black hole binaries), represent a key area of investigation in time-domain multi-messenger astronomy \citep[e.g.][]{Lloyd-Ronning2017,Lloyd-Ronning2022}.

Radio emission produced in the interaction of core-collapse SNe with the surrounding circumstellar matter (CSM) can provide a fast-forwarded view of the progenitor mass loss history starting from the moments right before explosion, thus shedding light on the still largely debated mechanisms driving mass loss in evolved massive stars \citep[e.g.][]{Corsi2014,Chandra2018,Stroh2021}.
Radio observations can also identify the rarest, radio-bright naked (Ib/c) core-collapse SNe, and clarify their yet-to-be-understood link to GRBs
\citep[ultra-relativistic explosions; e.g.][and references therein]{Corsi2022}.
The resolution provided by the ngVLA continental baselines could enlarge the currently limited sample of radio-emitting SN ejecta for which direct size measurements are possible \citep[e.g.][]{Bartel2017}.
These measurements represent the only clear way forward to disentangling unambiguously radio transients associated with an initial phase of relativistic expansion (off axis GRBs), from largely un-decelerated explosions (CSM-interacting radio SN).
New work has demonstrated how brief phases of super-Eddington accretion from a donor star onto a black hole or neutron star binary companion can power both luminous years-long transients, as well as millisecond Fast Radio Burst (FRB)-like emission \citep{Sridhar2022}.
This connects to the recognition that FRBs can be associated with persistent counterparts that are luminous enough to be seen at a distance of hundreds of Mpc \citep{Law2022,Dong2022}, and may comprise up to 1\% of the compact extragalactic sky.
The spatial resolution, sensitivity, and frequency coverage of the ngVLA will be crucial to identifying these radio sources, as they must be distinguished from nuclear AGN and are expected to have flat radio spectra.

Compact binary systems containing stellar mass black holes are golden targets for multi-messenger gravitational wave astronomy from the ground
\citep[e.g.][]{Abbott2017,Abbott2021b}.
With the planned sensitivity improvements of the LIGO, Virgo, and KAGRA detectors, predictions are that gravitational wave events localized to $\lesssim$ 100 deg$^{2}$ will keep growing by an order of magnitude in each of the LIGO-Virgo-KAGRA observing runs planned for the next 3-5 years  \citep{Petrov2022}.
Beyond that, and by the time the ngVLA becomes operational, a network of ground-based detectors of improved sensitivity will likely be providing gravitational wave triggers with substantially improved localization capabilities.
In this context, the ngVLA will be uniquely positioned to probe the diversity of jets with respect to their energy and speed distribution
\citep{Hallinan2017,Lazzati2018}, and in relation to the nature of their merger remnants \citep[e.g.][]{Radice2020,Nedora2021,Balasubramanian2022}, the physics of the interaction with the surrounding neutron-rich debris
\citep[e.g.][]{Kasliwal2017,Lazzati2019}, and the density of the local environments \citep{Fong2015}.
Polarization measurements combined with the exquisite ngVLA resolution could offer a unique opportunity to directly probe the structure of the jets and their magnetic fields \citep{Corsi2018,Mooley2018}.

It is important to stress that the radio follow up of gravitational waves may  go through a major transformation during the ngVLA lifetime.
Indeed,  next generation ground-based detectors such as  Cosmic Explorer \citep{Evans2021} and the Einstein Telescope
\citep{Maggiore2020} may come online in the 2040s.
These are facilities envisioned to supersede the existing LIGO and Virgo detectors with a factor of ten improvement in sensitivity.
As such, they are expected to push detections of compact binary systems to the peak of star formation, while providing several tens of events per year at $z leq$ 0.5 with sky localizations optimally matched to the field-of-view of the ngVLA at 2.4 GHz \citep{Borhanian2022}.
Hence, the ngVLA could shed light on the ejecta of black hole - neutron star systems for which radio observations may represent the only means to detect an EM counterpart to the gravitational wave event
\citep{Fragione2021}.

IMBHs will benefit from gravitational wave observations on the ground and in space, the lower end of the mass spectrum with ground-based gravitational wave observatories \citep[e.g.][and references therein]{Abbott2022}, and the upper mass range 10$^{2}$–10$^{4}$ M${_\odot}$ up to a redshift $z \sim$ 20 \citep{Saini2022} from space with LISA.
Modulation effects due to LISA’s orbital motion around the Sun facilitate precise premerger localization of the sources, which in turn would help in electromagnetic follow ups.
Electromagnetic emission from IMBH mergers may provide us vital clues about the environments where these mergers occur and the distance estimation can pave the way for cosmography.
The ngVLA, with its superb sensitivity, may play a critical role in this hunt for what are currently largely unknown electromagnetic counterparts of IMBH mergers.

The ngVLA will also be particularly important as a follow-up tool for the 10  - 100s of coalescing binary SMBHs detected by LISA each year.
A general prediction of General Relativistic MHD simulations of coalescing binary SMBHs is the occurrence of a collimated Poynting-flux outflow at the moment of coalescence, which may exceed the Eddington luminosity of the system.
Such an outflow is likely to be observable as a few-day to few-week to few-month radio transient peaking at $\nu \gg$ 10 GHz and exceeding 100 $\mu$Jy at a redshift $z \sim$ 6 \citep{Ravi2018,Mangiagli2022,Yuan2021}.

The generation of detectors following LISA will observe the tidal disruption and subsequent accretion of a star by a SMBH (Tidal Disruption Events, or TDEs) at cosmological distances, enabling population studies and constraints on the black hole mass function \citep{Pfister2022}.
Until that time, the ngVLA is the only planned instrument that can both discover and characterize a large number of these short-lived radio sources \citep{vanVelzen2018}.
Multi-frequency radio follow-up observations (3 - 100 GHz) of TDEs found in optical or X-ray surveys will provide a measurement of the jet efficiency as a function of black hole spin, thus enabling a direct test of the prediction that relativistic jets require high spin.
Hundreds of tidal disruption jets are expected to be discovered in a blind ngVLA survey for radio transients \citep{vanVelzen2018}.
Some of these sources could also be resolved via the long baselines of ngVLA, enabling a robust measurement of the jet launch date and the magnetic field strength.
Measurement of the accretion rate at the launch date proceeds from the thermal emission of the tidal disruption flare, and provides another unique opportunity to identify the conditions that lead to jet production.

High-energy cosmic neutrinos are the direct product of the most energetic cosmic particle accelerators.
The accretion of mass onto SMBHs at the center of their host galaxies is one of the most luminous persistent sources of electromagnetic radiation in the Universe, and are of interest as potentially powerful high-energy cosmic-ray accelerators \citep{Murase2017}.
Spatial coincidences with high-energy astrophysical neutrinos reveal blazars, the most powerful AGNs where jets are aligned with our line of sight, to be the likely association for the first few astrophysical neutrino detections \citep{IceCube2018,Rodrigues2021}.
This conclusion has been cemented by further analysis of IceCube data \citep{Plavin2022}, with the current understanding that a significant fraction of high-energy neutrinos might be produced in blazars in a region which is opaque to gamma-rays.
High resolution ngVLA observations probe beamed synchrotron radiation of parsec-scale jets, while neutrino emission undergoes similar beaming towards an observer.
The ngVLA observations would then provide unique information at the parsec and sub-parsec scales, and provide crucial information about proton acceleration to relativistic energies and consequent neutrino production.
SMBH coalescences, along with producing gravitational wave radiation, have strong accretion activities and powerful jets and are also promising candidates for neutrino counterpart emission originating from jet-induced shocks \citep{Murase2017}.
In these cases, the enhanced sensitivity, high-frequency capabilities,  and outstanding survey speed will be ideally suited to discover, localize, and unlock the scientific potential of multi-messenger transients  \citep{Mangiagli2020}.

\subsection{Telescope Requirements:}

The high-resolution capabilities of the ngVLA are crucial for surveying black holes.
High-resolution imaging will enable proper-motion separation of local black holes (both Galactic and in nearby galaxies, out to $\sim15$ Mpc) from background sources.
Milliarcsecond- and sub-milliarcsecond resolutions will enable the ngVLA to directly image and track the SMBH binaries that will be detected in gravitational waves by LISA and pulsar timing arrays
\citep[e.g.,][]{Wrobel2022}.
These astrometric science goals are feasible given the implementation of very long baselines ($\sim8000$ km for mas–$\mu$as accuracy).
Survey work will also require the ability to correlate on many phase-centers while using long baselines, so that positions of all sources can be measured in each epoch, and the astrometric selection of Galactic objects can be done simultaneously with the detection project.

The key frequency range is 5 - 20 GHz for most of this science.
The availability of higher frequencies will be critical for source follow-up in regions with high interstellar scatter broadening.
Extension to lower frequencies (down to 1-2 GHz) is necessary for pulsar timing array observations.

The ability to form sub-arrays operating independently with distinct frequency settings and/or on-sky positions is also crucial for this science.
Pulsar timing array observations require ideally at least 5 sub-arrays.
Rapid variability measurements of X-ray binaries and other objects require a large number of subarrays to ensure continuous coverage even while allowing for calibration (two sub-arrays would observe at each frequency to cover temporal gaps).
The ability to phase up the subarrays would increase the sensitivity of continuum observations of stochastic variability.

An essential requirement to enable time-domain and multi-messenger investigations is the ability to receive and respond to external triggers.
The triggering capability and high survey speed, covering a localization region to the appropriate depth, are absolutely critical to making the multi-messenger identifications which enable this breakthrough science.
In the case of both LIGO- and LISA- detected events, the localization of the event will be to roughly 10 square degrees at the time the ngVLA is operational.
The observing frequencies providing optimum detection of radio emission from NS-NS and NS-BH mergers identified from LIGO differ compared to those for coalescing SMBH binaries with LISA  (2.2 GHz with 2.7GHz bandwidth for the former, 27 GHz with 14 GHz bandwidth for the latter).
The ngVLA’s sensitivity enables mapping a 10 square degree region in less than 10 hours of on-the-fly mapping per epoch at both frequencies, leading to an optimal $\sim$ 1 $\mu$Jy detection level for NS-NS and NS-BH mergers, and $\sim$ 10 $\mu$Jy for SMBH mergers.

Critical aspects of the science of black holes described in the previous sections will be made unequivocally stronger by coordinated multi-wavelength observations (including but not limited to strictly simultaneous observations).
Once a candidate electromagnetic counterpart has been identified (as described above), follow-up observations with other facilities will be crucial to understanding the dynamic nature of the event.
This will require coordinated multi-wavelength observations and a mechanism by which requests for joint time at multiple facilities external to NRAO can be considered.
 Similarly, the ability to facilitate target-of-opportunity and director’s-time requests, potentially spanning multiple facilities, is crucial.

SMBH binaries discovered by pulsar timing arrays do not have such stringent requirements on mapping speed, given the slow evolution of the sources; a list of several thousand galaxies would require follow-up.
However, the ability to provide high-resolution imaging for several hundred to several thousand galaxies, to a depth that would enable detection of a Low Luminosity AGN (LLAGN), and at several epochs covering multiple orbital periods (binary orbits of weeks to years) are required.

\subsection{Data Product and Processing Requirements:   }

The minimum resource case for high-level data products to be delivered and archived for this science goal includes both raw and calibrated uv data.
In addition, science-ready multi-frequency synthesis images out to the FWHM of the array will suffice, with Taylor-term imaging to account for the large relative bandwidths.

The ideal case for high-level data products encompasses raw and calibrated data, but includes the option for self-calibration or not (self-calibration is not appropriate for the astrometric uses described above). The production of science-ready multi-frequency synthesis images out to two times the FWHM of the array including the first Taylor term would be ideal. In addition, reporting of some basic data quality checks on averaged and time-resolved spectra, including percent of data flagged, would add information to enable the scientist to assess the utility of the data in answering the science goals. Some additional information determined from science-ready images are: a list of sources above 6 sigma (centroid position, integrated and peak flux density, source shape); image statistics like beam size, peak flux, standard deviation, and mean; and a list of known in-beam calibrators.Time-series data products that probe variability within an observation would also be desirable to accomplish these science goals. For the astrometric goals of separating black hole candidates from background or foreground objects, the ability to map a field at low resolution and then do a second correlator pass to do astrometry using the long baselines would provide a major telescope time cost savings for wide field astrometric surveys of the Galactic Plane.

Ideally high-level data products should be both standardized for archive usage and tuned to the PI proposal request. Visibility processing to complete this key science goal should include the ability to create images as a function of time and spectral bin, as well as cut along different spectral windows and times. The latter is needed to track astrophysical sources  whose signal drifts in frequency and time, either due to  internally-generated mechanisms (e.g. plasma motions) or propagation effects. For AGN and SMBHB, for instance, such a fine-tuned data cut would enable searches for spectral curvature.  To maximize the additional science which can be performed on this data by archival researchers, archiving as raw a data product as possible will be critical; this assumes that new science or new techniques require the download of relatively unprocessed archival data products. No highly specialized products beyond those described above are foreseen to fully exploit the data from an archive perspective.

There is no strong or compelling use case for single node deployment for running ngCASA.

An additional capability in ngVLA-provided data analysis software suite  beyond CASA tasks, the CASA viewer, and/or CARTA includes a standardized source extractor performed on all images produced from science ready data products, to maximize the science from high level data products.

\bigskip\bigskip\bigskip

\color{black}
\paragraph*{Acknowledgements}
The ngVLA Science Advisory Council thanks past members and the many in the international community 
who contributed to the original development and formulation of ngVLA Key Science Goals, as well as
those who engaged in the process of this revision. 
The National Radio Astronomy Observatory is a facility of the National Science Foundation operated 
under cooperative agreement by Associated Universities, Inc.


\clearpage

\bibliography{ksg1,ksg2-bam,ksg3,ksg4,ksg5,ksg2-jbb,appendix.bib}
\bibliographystyle{aasjournal}

\clearpage

\appendix

\section{Summary of Data Processing Requirements based on ngVLA Key Science Goals}

The ngVLA document {\it Observing Modes Calibration Strategy} \citep{observing_modes_calibration_strategy}
outlines proposed ngVLA observing modes and the data products that will be created for the envisioned science use cases, 
not limited to the KSGs. In this framework, there will be seven fundamental sets of data products: 
\begin{itemize}
    \item{interferometric continuum imaging (IQUV) with Stokes I spectral cubes;}
    \item{interferometric spectral polarization imaging (Stokes IQUV);}
    \item{pulsar search mode data (power vs. time/frequency/polarization);}
    \item{pulsar timing mode data (folded pulse profiles);}
    \item{phased beam baseband channels (VLBI or radar recording);}
    \item{single-dish autocorrelation imaging (analogous to single-dish mapping); and}
    \item{single-dish /interferometry combined imaging (continuum and cubes).}
\end{itemize}

These standard data products will be sufficient to achieve many of the objectives of the Key Science Goals.
However, they are by no means exhaustive, and other products and processing are recognized to be potentially valuable 
and possibly essential. These include per execution or per epoch images,  spectral index images, rotation measure cubes, 
calibrated visibilities,  autocorrelation spectra, time-critical processing, processing of multiple phase centers, 
options for reprocessing and self-calibration, and higher order products (e.g., source finder, spectral line finder).
Table~\ref{tab:dataproducts} summarizes additional requirements and desirable data product/processing capabilities 
for achieving the Key Science Goals. 

Not yet addressed in this document is the discussion of the merits of a science platform, or a common, potentially cloud, computing platform available for all scientists to work on ngVLA data. The data volumes of the ngVLA will eclipse the current VLA significantly (likely by factors of $10^3$), and so there are compelling reasons to consider implementing a science platform model. A science platform (such as the CANFAR model\footnote{www.canfar.net} which has been presented to the Science Advisory Council) would provide: equity of access to resources for researchers to take data from archive to publication; mitigation of the environmental impact of redundant data downloads and computing effort by multiple team members in the absence of a common work area; and enhanced team access to data, processing tools and final data products. A science platform will be particularly advantageous for large Legacy style projects and the KSGs, which are likely to have broad participation across the ngVLA user base, which is global. Accessible technology will be a key component to the success of these projects. 

\begin{table}[h!]
\centering
\begin{tabular}{| l |  c |  c |  c |  c |  c |} 
\hline
~ & KSG1 & KSG2 & KSG3 & KSG4 & KSG5 \\
{\bf Visibility data requirements} & ~ & ~ & ~ & ~ & ~ \\
\hline
Raw and calibrated visibilities, 
pre-averaged in time and frequency as appropriate 
  & X & X & X & ~ & X \\
Option to reprocess visibility data for 
weighting/imaging variations  
  & X & X & X & ~ & ~ \\
Preservation of native time and channel spacing 
weighting/imaging variations  
  & X & X & X & ~ & ~ \\
Maximum bandwidth at full spectral resolution 
  & ~ & X & ~ & ~ & ~  \\
Ability to target spectral lines with (many, i.e., dozens) 
individual high spectral resolution windows 
  & X & X & X & ~ & ~  \\
Robust tools to combine main array data with
short baseline array and single dish data
  & ~ & X & X & ~ & ~ \\
Access to beamforned pulsar data 
  & ~ & ~ & ~ & X & ~ \\
PSRFITS flux- and polarization-calibrated profiles 
  & ~ & ~ & ~ & X & ~ \\
RFI cleaning of PSRFITS profiles 
  & ~ & ~ & ~ & X & ~ \\
Pulse phase-binned (gated) visibility data
  & ~ & ~ & ~ & X & ~ \\
  
~ & ~ & ~ & ~ & ~ & ~ \\
{\bf Imaging and spectra} & ~ & ~ & ~ & ~ & ~ \\
Science-ready multi-frequency synthesis images to $2-3\times$ FWHM
  & X & ~ & ~ & ~ & X \\
Option for self-calibration 
  & X & X & X & ~ & ~ \\
Basic data quality checks on averaged/time-resolved spectra
  & X & ~ & ~ & ~ & X \\
Images as a function of time and spectral bin 
  & ~ & ~ & ~ & ~ & X \\
Images as a function of pulse phase bin
  & ~ & ~ & ~ & X & ~ \\
  
~ & ~ & ~ & ~ & ~ & ~ \\
{\bf Catalogs} & ~ & ~ & ~ & ~ & ~ \\
Standardized source extraction on all images
  & ~ & ~ & ~ & ~ & X \\
list of sources above $6\sigma$ (position, flux density, shape)
  & ~ & ~ & ~ & ~ & X \\
image statistics (beam size, peak flux, rms, known in-beam calibrators)
  & ~ & ~ & ~ & ~ & X \\
  
~ & ~ & ~ & ~ & ~ & ~ \\
{\bf Other data products } & ~ & ~ & ~ & ~ & ~ \\
Time series data products to probe variability within an observation 
  & ~ & ~ & ~ & ~ & X \\
  
~ & ~ & ~ & ~ & ~ & ~ \\
{\bf Specialized Software/Processing} & ~ & ~ & ~ & ~ & ~ \\
Standard pulsar search/calibration/timing software and pipelines 
  & ~ & ~ & ~ & X & ~ \\
Pipeline for pulsar fold-mode data products (simmilar to NANOGrav)
  & ~ & ~ & ~ & X & ~ \\
Stacking routines 
  & ~ & X & ~ & ~ & ~ \\
  
~ & ~ & ~ & ~ & ~ & ~ \\
{\bf Other requirements} & ~ & ~ & ~ & ~ & ~ \\
Adequate storage space for pulsar raw search-mode data 
  & ~ & ~ & ~ & X & ~ \\
Standardized data products for archive usage and tuned to PI request
  & ~ & X & ~ & ~ & X \\
Extra processing requirements for large-area mosaics
  & ~ & ~ & X & ~ & ~ \\
Tracking of RFI 
  & ~ & ~ & ~ & X & ~ \\
Ability to image at low resolution followed by long baselines for astrometry
  & ~ & ~ & ~ & ~ & X \\
Archiving as raw a data product as possible
  & X & ~ & ~ & ~ & X \\
  
\hline
\hline
\hline
\end{tabular}
\caption{\em 
Desirable data products and data processing capabilities based on the ngVLA Key Science Goals.
}
\label{tab:dataproducts}
\end{table}

\bigskip\bigskip\bigskip


\end{document}